\documentclass[twocolumn,pra,showpacs]{revtex4}
\usepackage{amssymb}
\usepackage{amsfonts}
\usepackage{amsmath}
\usepackage{graphicx}

\input{tcilatex}
\begin{document}

\title{Open quantum system approach to single-molecule spectroscopy}
\author{Adri\'{a}n A. Budini}
\affiliation{Consejo Nacional de Investigaciones Cient\'{\i}ficas y T\'{e}cnicas, Centro
At\'{o}mico Bariloche, Av. E. Bustillo Km 9.5, (8400) Bariloche, Argentina,
and Consortium of the Americas for Interdisciplinary Science and Department
of Physics and Astronomy, University of New Mexico, Albuquerque, New Mexico
87131, USA}
\date{\today}

\begin{abstract}
In this paper, single-molecule spectroscopy experiments based on continuous
laser excitation are characterized through an open quantum system approach.
The evolution of the fluorophore system follows from an effective
Hamiltonian microscopic dynamic where its characteristic parameters, i.e.,
its electric dipole, transition frequency, and Rabi frequency, as well as
the quantization of the background electromagnetic field and their mutual
interaction, are defined in an extended Hilbert space associated to the
different configurational states of the local nano-environment. After
tracing out the electromagnetic field and the configurational states, the
fluorophore density matrix is written in terms of a Lindblad rate equation.
Observables associated to the scattered laser field, like optical spectrum,
intensity-intensity correlation, and photon-counting statistics, are
obtained from a quantum-electrodynamic calculation also based on the
effective microscopic dynamic. In contrast with stochastic models, this
approach allows to describe in a unified way both the full quantum nature of
the scattered laser field as well as the classical nature of the environment
fluctuations. By analyzing different processes such as spectral diffusion,
lifetime fluctuations, and light assisted processes, we exemplify the power
of the present approach.
\end{abstract}

\maketitle

\section{Introduction}

In spite that in recent years many different experimental techniques have
been established, single molecule spectroscopy (SMS) \cite%
{barkaiChem,michel,jung}, that is, the study of single nano-objects is
predominantly realized by purely optical methods, i.e., by measuring the
far-electromagnetic field scattered by the system when it is subjected to
laser radiation.

In contrast to standard quantum optical systems \cite%
{breuerbook,carmichaelbook,milburn,scully,loudon,yariv}, where the reservoir
is only defined by the background (free) quantized electromagnetic field, in
SMS the supporting nano-environment is highly structured. Its definition and
specific properties depend on each experimental setup. In fact, the local
environment felt by the fluorophore may involve cryogenic single-molecules 
\cite{tamarat,donley,kulzer}, molecules at room temperature, either
immobilized \cite{trautman,MoernerOrrit} or diffusing in solution \cite%
{Elson}, the solid state matrix supporting a nanocrystal \cite%
{nirmal,PMichler,brokmann,verberk,kuno,potatova,cichosLaser,cichos,vanOrden,bianco}%
, or single bio-molecules \cite%
{Dickson,xie,WEMoerner,caoYang,Lu,SWeiss,valle}. Even more, the proper
definition of the background electromagnetic field may be modified because
the local surrounding of the system may develop fluctuations in its
dielectric properties \cite{cichos,valle}.

Each specific environment leads to different underlying physical or chemical
processes that in turn modify the emission properties of the fluorophore. In
most of the situations, it can be model as a two-level optical transition.
The central theoretical problem of SMS is to relate the scattered laser
field statistics with the underlying nanoscopic environment dynamic. Due to
its complexity, its dynamical influence is usually taken into account by
adding random elements to the characteristic parameters of the system \cite%
{barkaiChem,michel,jung}. For example, spectral diffusion processes are
taken into account by adding classical noise fluctuations to the transition
frequency of the system, while lifetime fluctuations may be associated to
transitions between different conformational (chemical or physical) states
of the environment. Stochastic Bloch equations \cite%
{zheng,he,FLHBrown,BarkaiExactMandel}, stochastic decay rates \cite%
{wang,brown}, modulated reaction models \cite{caoYang,xieSchenter} or
stochastic reaction coordinates \cite{mukamel} are some of the associated
theoretical models. On the other hand, radiation patterns whose statistics
depend on the external laser power \cite%
{Dickson,WEMoerner,hoch,barbara,talon, moerner,wild,cichosLaser} are modeled
by introducing extra states coupled incoherently to the upper level of the
system \cite{molski}.

The previous approaches are well-accepted and standard theoretical tools for
modeling SMS experiments. They provide a solid basis for describing
different\ experimental observables, such as those associated to the photon
counting statistics. Nevertheless, as the influence of both the background
electromagnetic field and the reservoir fluctuations are represented in a
unified way by a classical noise, in general it is not clear how to
calculate arbitrary observables associated to the scattered laser field.
This limitation is evident when considering, for example, the optical or
absorption spectrum, which is defined in terms of the scattered field
correlations.

The previous drawback or limitation could be surpassed if one is able to
describe SMS experiments on the basis of a full open quantum system
approach. By an open quantum systems approach we mean: (i) the possibility
of describing both the fluorophore system and the reservoir fluctuations
through a density matrix operator, whose evolution, i.e., its master
equation, can be written and adapted to each specific situation. (ii)
Describing the quantum nature of the scattered electromagnetic field through
operators and providing a closed solution for their evolution, in such a way
that a simple procedure for calculating arbitrary field correlations is
established. Then, the calculation of observables like the optical spectrum
follows straightforwardly. (iii) Characterizing the photon counting process
through a Mandel formula and establishing a manageable analytical tool for
the explicit calculation of the photon counting probabilities. The main goal
of this paper is to demonstrate that it is possible to build up such kind of
powerful and general approach, which not only recovers the predictions of
standard approaches, but also allows to characterize, in different
situations, arbitrary observables associated to the scattered quantum
electromagnetic field.

The scarce application of an open quantum system approach to SMS has a clear
origin. Due to the complexity of the underlying nano-environment, a
microscopic description, from where to deduce the system density matrix
evolution, is lacking. We overcome this difficulty by noting that the noise
fluctuations induced by the reservoir \cite{barkaiChem,michel,jung} can be
associated to a coarse grained representation of its complex structure \cite%
{vanKampen}, allowing us to write microscopic interactions that take into
account its leading dynamical effects and at the same time are analytically
manageable. From the effective microscopic dynamic the system and reservoir
fluctuations result described in terms of a Lindblad rate equation. The
theoretical validity of these equations for describing non-Markovian open
quantum system dynamics was established in Refs. \cite{rate,breuer}. The
possibility of establishing a quantum-electrodynamic treatment of SMS
experiments also relies on the results of Ref. \cite{QRT}, where the quantum
regression hypothesis was analyzed for Lindblad rate master equations.

We remark that recent author contributions anticipated the possibility of
establishing the powerful formalism developed in this paper. In Refs. \cite%
{rapid}, it was demonstrated that an anomalous fluorescence blinking
phenomenon can be dynamically induced by a complex environment whose action
can be described by a direct sum of Markovian sub-reservoirs. In Ref. \cite%
{mandel}, for the same situation, observables associated to the photon
counting process and a mapping with triplet blinking models \cite%
{barkaiChem,molski} were characterized. In Ref. \cite{luzAssisted}, we shown
that fluorescence blinking patterns whose statistics depend on the external
laser power can be model through an underlying tripartite interaction. Both
situations lead to quantum master equations that correspond to particular
cases of the present general formalism. Since in both cases the calculations
relied on specific generalizations of a quantum jump approach \cite{plenio},
there is not a recipe for calculating the scattered field correlations. The
present quantum-electrodynamic treatment also fills up this gap.

The paper is outlined as follows. In Sec. II, the effective Hamiltonian
microscopic dynamic that defines the full approach is introduced and the
evolution of the fluorophore density matrix is obtained. In Sec. III,
observables associated to the scattered laser field are derived from the
full microscopic approach. In Sec. IV, the photon counting statistics is
characterized through a Mandel formula and a generating function approach 
\cite{cook,mukamelPRA}. In Sec. V, the formalism is exemplified by analyzing
different kind of environment fluctuations. Processes like spectral
diffusion, lifetime fluctuations, and light assisted processes are
explicitly characterized through the scattered field observables. In Sec. VI
we give the conclusions.

\section{Effective Microscopic description and density matrix evolution}

The total Hilbert space $\mathcal{H}_{\mathcal{T}}$ associated to a SMS
experiment is defined by the external product $\mathcal{H}_{\mathcal{T}}=%
\mathcal{H}_{\mathcal{S}}\otimes \mathcal{H}_{\mathcal{F}}\otimes \mathcal{H}%
_{\mathcal{B}},$ where each contribution denotes respectively the Hilbert
space of: the fluorophore system, the background electromagnetic field, and
the rest of the degrees of freedom that define the local nano-reservoir felt
by the system. In general, it is impossible to know the total microscopic
dynamic of the reservoir $\mathcal{B}$. In order to bypass this task,
following an argument presented by van Kampen \cite{vanKampen}, we notice
that the complexity of the reservoir may admits a simpler general
description. Since the nano-reservoir can only be indirectly observed
through the fluorophore system, its Hilbert space structure can not be
resolved beyond the experimental resolution. Therefore, it is split as $%
\mathcal{H}_{\mathcal{B}}=\oplus _{R=1}^{R_{\max }}\mathcal{H}_{\mathcal{B}%
_{R}},$ where each subspace $\mathcal{H}_{\mathcal{B}_{R}}$ is defined by
the set of all quantum states that lead to the same system dynamic \cite%
{vanKampen}. As the reservoir may be characterized by inordinately dense as
well as by discrete manifolds of energy levels, some $\mathcal{H}_{\mathcal{B%
}_{R}}$ may be of finite dimension.

Clearly, the subspaces $\{\mathcal{H}_{\mathcal{B}_{R}}\}_{R=1}^{R_{max}}$
define the maximal information about the reservoir Hilbert space structure
that can be achieved from a SMS experiment. Then, our approach consists in
representing the reservoir through a set of coarse grained \textquotedblleft
configurational macrostates\textquotedblright $\{|R\rangle \}_{R=1}^{R_{\max
}}$, $\langle R|R^{\prime }\rangle =\delta _{RR^{\prime }}$, each one being
associated to each subspace, and writing the total microscopic dynamic in
the effective Hilbert space $\mathcal{H}_{\mathcal{T}}^{eff}=\mathcal{H}_{%
\mathcal{S}}\otimes \mathcal{H}_{\mathcal{F}}\otimes \mathcal{H}_{\mathcal{U}%
}.$ The configurational Hilbert space $\mathcal{H}_{\mathcal{U}}$ is
expanded by the (unknown) configurational macrostates. The approach is
closed after defining both, the field quantization and its interaction with
the system in the total effective Hilbert space, and the self-dynamics of
the configurational states.

\subsection{Electromagnetic field quantization}

The classical Maxwell equations \cite{landau} 
\begin{subequations}
\begin{eqnarray}
\mathbf{\bigtriangledown }\cdot \mathbf{D} &=&0,\ \ \ \ \ \ \ \mathbf{%
\bigtriangledown }\times \mathbf{E}=-\frac{1}{c}\frac{\partial \mathbf{B}}{%
\partial t}, \\
\mathbf{\bigtriangledown }\cdot \mathbf{B} &=&0,\ \ \ \ \ \ \ \mathbf{%
\bigtriangledown }\times \mathbf{H}=+\frac{1}{c}\frac{\partial \mathbf{D}}{%
\partial t},
\end{eqnarray}%
provide the basis for the quantization of the electromagnetic field in a
material media \cite{loudon,yariv}. The macroscopic field vectors are
related by the constitutive relations $\mathbf{D}=\varepsilon \mathbf{E},$
and $\mathbf{B}=\mu \mathbf{H},$ where $\varepsilon $ and $\mu $\ are the
macroscopic material constants and $c$ is the light velocity in free space.
For an absorptionless dielectric, the Maxwell equations lead to the
classical Hamiltonian field 
\end{subequations}
\begin{equation}
H_{\mathcal{F}}=\frac{1}{8\pi }\int_{V}(\varepsilon |\mathbf{E}|^{2}\mathbf{+%
}\mu |\mathbf{H}|^{2})dV,  \label{ClassicalFieldHamiltonian}
\end{equation}%
where $V$ is the volume of integration.

The canonical field quantization follows from the expansion of the electric
and magnetic field in normal modes and by associating to each component a
creation and annihilation photon operator. Here, the domain of quantization $%
V$ is defined by the local surrounding of the system and not by the full
volume of the supporting media. In order to take into account the possible
dependence of the local field on the configurational bath states, the
macroscopic material constants associated to the volume $V$ are written as
operators in $\mathcal{H}_{\mathcal{U}},$%
\begin{equation}
\varepsilon =\sum_{R}\varepsilon _{R}\left\vert R\right\rangle \left\langle
R\right\vert ,\ \ \ \ \ \ \ \ \ \ \mu =\sum_{R}\mu _{R}\left\vert
R\right\rangle \left\langle R\right\vert .
\end{equation}%
Then, depending on the environment configurational state $\{\left\vert
R\right\rangle \},$ the local field is quantized in a media characterized by
the (real) constants $\{\varepsilon _{R},\mu _{R}\}.$ Consistently, as the
system have nanoscopic dimensions we assume that these constants, in the
local surrounding of the fluorophore $(V),$ do not have any spatial
dependence. After standard calculations \cite{carmichaelbook}, both the
electric $\mathbf{E}=\mathbf{E}^{(+)}+\mathbf{E}^{(-)}$ and magnetic $%
\mathbf{H}=\mathbf{H}^{(+)}+\mathbf{H}^{(-)}$ field operators can be written
in terms of positive and negative frequency contributions, each one related
to the other by an Hermitian conjugation operation. They read 
\begin{subequations}
\label{Campos}
\begin{eqnarray}
\mathbf{E}^{(+)}\! &=&\!i\sum_{R}\left\vert R\right\rangle \left\langle
R\right\vert \sum_{\mathbf{k},\lambda }\!\Big(\frac{\hbar \omega _{\mathbf{k}%
}^{(R)}}{2V\varepsilon _{R}}\Big)^{1/2}\!\hat{e}_{\mathbf{k}\lambda }e^{i%
\mathbf{k\cdot r}}a_{\mathbf{k}\lambda },\ \ \ \ \   \label{Electric} \\
\mathbf{H}^{(+)}\! &=&\!i\sum_{R}\left\vert R\right\rangle \left\langle
R\right\vert \sum_{\mathbf{k},\lambda }\!\Big(\frac{\hbar \omega _{\mathbf{k}%
}^{(R)}}{2V\mu _{R}}\Big)^{1/2}\!\hat{e}_{\mathbf{k}\lambda }^{\prime }e^{i%
\mathbf{k\cdot r}}a_{\mathbf{k}\lambda }.\ \ \ \   \label{Magnetic}
\end{eqnarray}%
Here, we assumed that inside the volume $V$ the normal modes can be well
approximated by plane waves. As usual, the creation and annihilation
operators satisfy the commutation relation $[a_{\mathbf{k}\lambda },a_{%
\mathbf{k}^{\prime }\lambda ^{\prime }}^{\dag }]=\delta _{\mathbf{kk}%
^{\prime }}\delta _{\lambda \lambda ^{\prime }},$ where $\mathbf{k}$ and $%
\lambda $ denote the wave vector and polarization of the quantized mode
respectively. In agreement with the transversability of the electromagnetic
field, the polarization of the magnetic field reads $\hat{e}_{\mathbf{k}%
\lambda }^{\prime }=(\mathbf{\hat{k}\times }\hat{e}_{\mathbf{k}\lambda }),$
where $\mathbf{\hat{k}}\equiv \mathbf{k/|k|}.$ $\mathbf{r}$ denotes the
position vector. The dispersion relation associated to each configurational
state is 
\end{subequations}
\begin{equation}
\omega _{\mathbf{k}}^{(R)}=\frac{c}{\sqrt{\varepsilon _{R}\mu _{R}}}|\mathbf{%
k}|.  \label{dispersion}
\end{equation}%
From now on, we take $\mu _{R}\rightarrow 1.$ As is well known \cite{landau}%
, this assumption is always valid in an optical regime.

\subsection{System and field Hamiltonians}

The fluorophore system is defined by a two-level optical transition with
natural frequency $\omega _{A}.$ The upper and lower states are denoted as $%
\left\vert b\right\rangle $ and $\left\vert a\right\rangle $ respectively.
We take into account that, depending on the configurational state $%
\left\vert R\right\rangle ,$ the natural frequency $\omega _{A}$ may be
shifted a quantity $\delta \omega _{A}^{(R)}.$ These parameters define the
spectral shift of the system associated to each state of the reservoir \cite%
{barkaiChem}. Therefore, the system and field Hamiltonians are written as 
\begin{equation}
H_{\mathcal{S}}=\hbar \omega _{0}\sigma _{z}/2,\ \ \ \ \ \ \ \ H_{\mathcal{F}%
}=\sum\nolimits_{\mathbf{k}\lambda }\hbar \omega _{\mathbf{k}\lambda }a_{%
\mathbf{k}\lambda }^{\dag }a_{\mathbf{k}\lambda },  \label{Hfield}
\end{equation}%
where $H_{\mathcal{F}}$ follows from Eqs.~(\ref{Campos}) and (\ref%
{ClassicalFieldHamiltonian}). $\sigma _{z}$ is the z-Pauli matrix written in
the base $\{\left\vert a\right\rangle ,\left\vert b\right\rangle \}.$ The
frequency operator of the system is defined by%
\begin{equation}
\omega _{0}=\omega _{A}+\sum_{R}\delta \omega _{A}^{(R)}\left\vert
R\right\rangle \left\langle R\right\vert ,  \label{Omega_Cero}
\end{equation}%
while for the field it reads%
\begin{equation}
\omega _{\mathbf{k}\lambda }=\sum_{R}\omega _{\mathbf{k}\lambda
}^{(R)}\left\vert R\right\rangle \left\langle R\right\vert .
\label{OmegaBath}
\end{equation}%
The frequencies $\omega _{\mathbf{k}\lambda }^{(R)}$ are defined by the
dispersion relations Eq.~(\ref{dispersion}).

\subsection{Dipole-field interaction}

The natural decay of the fluorophore is induced by the coupling of its
electric dipole with the quantized electric field. Their interaction can be
written as%
\begin{equation}
H_{\mathrm{dip}}=-\mathbf{E}_{\mathbf{r}_{\mathbf{A}}}\cdot \mathbf{d}_{%
\mathbf{A}},  \label{Hdipolar}
\end{equation}%
where $\mathbf{d}_{\mathbf{E}}$ is the (electric) dipole operator associated
to the optical transition and $\mathbf{E}_{\mathbf{r}_{\mathbf{A}}}$ is the
electric field operator at the position $\mathbf{r}_{\mathbf{A}}$ of the
system. To take into account the different configurational states of the
reservoir, the standard definition of the dipole operator $\mathbf{d}_{%
\mathbf{A}}$ \cite{carmichaelbook}\ is generalized as 
\begin{equation}
\mathbf{d}_{\mathbf{A}}=\sum_{R,R^{\prime }}\mathbf{d}_{RR^{\prime }}(\sigma
^{\dag }\left\vert R\right\rangle \left\langle R^{\prime }\right\vert
+\sigma \left\vert R^{\prime }\right\rangle \left\langle R\right\vert ),
\label{DipoloOperator}
\end{equation}%
where $\{\mathbf{d}_{RR^{\prime }}\}$ are vectors (assumed real) with units
of electric dipole. $\sigma ^{\dag }$ and $\sigma $ are respectively the
raising and lowering operators acting on the states $\{\left\vert
a\right\rangle ,\left\vert b\right\rangle \}.$ The diagonal contributions $%
\mathbf{d}_{RR}$ define the dipole associated to each configurational state.
The nondiagonal contributions, $\mathbf{d}_{RR^{\prime }}$ with $R\neq
R^{\prime },$ take into account the possibility of coupling different
configurational states (those of finite dimension) through system
transitions.

The dipole-field interaction Eq.~(\ref{Hdipolar}), in a rotating wave
approximation \cite{carmichaelbook,loudon,milburn,scully}, from Eqs.~(\ref%
{Campos}) and (\ref{DipoloOperator}) reads%
\begin{equation}
H_{\mathrm{dip}}=\sum_{\substack{ \mathbf{k},\lambda  \\ R,R^{\prime }}}%
(\kappa _{RR^{\prime }}^{\mathbf{k},\lambda }\sigma ^{\dag }\left\vert
R\right\rangle \left\langle R^{\prime }\right\vert a_{\mathbf{k}\lambda
}+\kappa _{RR^{\prime }}^{\mathbf{k},\lambda ^{\ast }}\sigma \left\vert
R^{\prime }\right\rangle \left\langle R\right\vert a_{\mathbf{k}\lambda
}^{\dag }),  \label{HSUB}
\end{equation}%
where the interaction parameters are defined by%
\begin{equation}
\kappa _{RR^{\prime }}^{\mathbf{k},\lambda }=-i\Big(\frac{\hbar \omega _{%
\mathbf{k}}^{(R)}}{2V\varepsilon _{R}}\Big)^{1/2}e^{i\mathbf{k\cdot \mathbf{r%
}_{A}}}(\hat{e}_{\mathbf{k}\lambda }\cdot \mathbf{d}_{RR^{\prime }}).
\label{Acoples}
\end{equation}

\subsection{Environment configurational fluctuations}

The previous definitions effectively take into account the influence of the
different configurational states of the environment on the fluorophore and
the background electromagnetic field as well as into their mutual
interaction. Now, in agreement with the van Kampen argument \cite{vanKampen}%
, and consistently with real specific situations \cite%
{barkaiChem,michel,jung}, where the properties of the nano-environment
fluctuates in time, the configurational states (associated to dense bath
manifolds) must be endowed with a mechanism able to induce transitions
between them. We remark that the results developed in Refs. \cite%
{rapid,mandel,luzAssisted} do not take into account neither rely on this
extra dynamical effect, which in the context of an open quantum system
approach can only be recover with the present treatment.

In order to maintain a full microscopic description of the effective
dynamics, here the (incoherent) transitions $\left\vert R\right\rangle
\leftrightarrow \left\vert R^{\prime }\right\rangle $ are introduced through
the Hamiltonian 
\begin{equation}
H_{\mathcal{U}}^{\prime }=H_{\mathcal{U}}+H_{\mathcal{W}}+H_{\mathcal{UW}}.
\label{U_Hamiltonian}
\end{equation}%
$H_{\mathcal{U}}$ is the free Hamiltonian of $\mathcal{U},$ $H_{\mathcal{W}}$
define an environment ($\mathcal{W}$) responsible for the transitions $%
\left\vert R\right\rangle \leftrightarrow \left\vert R^{\prime
}\right\rangle ,$ while $H_{\mathcal{UW}}$ defines their mutual interaction.
We assume that the states $\{\left\vert R\right\rangle \}$ are the
eigenvalues of $H_{\mathcal{U}}$ with eigenvalues $\hbar \omega _{R},$ and $%
\mathcal{W}$ is defined by a continuous set of arbitrary bosonic normal modes%
\begin{equation}
H_{\mathcal{U}}=\sum\nolimits_{R}\hbar \omega _{R}\left\vert R\right\rangle
\left\langle R\right\vert ,\ \ \ \ \ \ \ H_{\mathcal{W}}=\sum\nolimits_{j}%
\hbar \omega _{j}b_{j}^{\dag }b_{j}.
\end{equation}
$b_{j}^{\dag }$ and $b_{j}$ are the \ creation and annihilation operators of 
$\mathcal{W}$ respectively. The Hamiltonian $H_{\mathcal{UW}}$ reads%
\begin{equation}
H_{\mathcal{UW}}=\hbar \sum\limits_{\substack{ j,R,R^{\prime }  \\ R\neq
R^{\prime }}}(\chi _{RR^{\prime }}^{j}\left\vert R\right\rangle \left\langle
R^{\prime }\right\vert b_{j}+\chi _{RR^{\prime }}^{j^{\ast }}\left\vert
R^{\prime }\right\rangle \left\langle R\right\vert b_{j}^{\dag }).
\label{Huw}
\end{equation}%
Then, the transitions $\left\vert R\right\rangle \leftrightarrow \left\vert
R^{\prime }\right\rangle $ are assisted by the creation or destruction of
bosonic excitations in each of the modes of frequency $\omega _{j}.$

\subsection{System density matrix evolution}

The previous analysis allow to define the unitary dynamic, $(d/dt)\rho _{%
\mathcal{T}}(t)=-(i/\hbar )[H_{\mathcal{T}},\rho _{\mathcal{T}}(t)],$ of the
density matrix $\rho _{\mathcal{T}}(t)$ associated to the Hilbert space $%
\mathcal{H}_{\mathcal{T}}^{eff}=\mathcal{H}_{\mathcal{S}}\otimes \mathcal{H}%
_{\mathcal{F}}\otimes \mathcal{H}_{\mathcal{U}}.$ The total Hamiltonian read 
\begin{equation}
H_{\mathcal{T}}=H_{\mathcal{S}}+H_{\mathcal{U}}^{\prime }+H_{\mathcal{F}}+H_{%
\mathrm{dip}},  \label{Htotal}
\end{equation}%
where each contribution follows from Eqs.~(\ref{Hfield}), (\ref{HSUB}) and (%
\ref{U_Hamiltonian}). $\rho _{\mathcal{T}}(t)$ describe the statistical
dynamical behavior of the system, the electromagnetic field, and the
configurational states. The joint dynamic of the fluorophore and the
configurational states is encoded in the density matrix $\rho _{\mathcal{SU}%
}(t),$ which follows after tracing out the degrees of freedom of the
background electromagnetic field $\mathcal{F}$\ and the reservoir $\mathcal{W%
},$ i.e., $\rho _{\mathcal{SU}}(t)=\mathrm{Tr}_{\mathcal{FW}}[\rho _{%
\mathcal{T}}(t)].$ The system density matrix $\rho _{\mathcal{S}}(t)=\mathrm{%
Tr}_{\mathcal{U}}[\rho _{\mathcal{SU}}(t)],$ can always be written as%
\begin{equation}
\rho _{\mathcal{S}}(t)=\sum\nolimits_{R}\rho _{R}(t),  \label{RhoS}
\end{equation}%
where the base $\{\left\vert R\right\rangle \}$ was used for taking the
trace over $\mathcal{U}.$ The auxiliary states $\rho _{R}(t)\equiv
\left\langle R\right\vert \rho _{\mathcal{SU}}(t)\left\vert R\right\rangle ,$
define the system dynamic \textquotedblleft \textit{given}%
\textquotedblright\ that the reservoir is in the configurational state $%
\left\vert R\right\rangle .$ The dynamic of the configurational states
follows from the populations%
\begin{equation}
P_{R}(t)=\mathrm{Tr}_{\mathcal{S}}[\rho _{R}(t)],
\label{ConfigurationalPopulation}
\end{equation}%
which provide the probability that the reservoir is in the configurational
state $\left\vert R\right\rangle $ at time $t.$ Therefore, these objects
encode the statistical properties of the noise fluctuations introduced in
standard stochastic approaches \cite{barkaiChem,michel,jung}. The expression
Eq.~(\ref{ConfigurationalPopulation}) follows straightforwardly from the
definition $P_{R}(t)\equiv \left\langle R\right\vert \rho _{\mathcal{U}%
}(t)\left\vert R\right\rangle ,$ where $\rho _{\mathcal{U}}(t)=\mathrm{Tr}_{%
\mathcal{S}}[\rho _{\mathcal{SU}}(t)].$

Both, the electromagnetic field $\mathcal{F}$\ and the bath $\mathcal{W}$
are assumed to be Markovian reservoirs. Then, their influence can be
described through a standard Born-Markov approximation. After some algebra 
\cite{carmichaelbook}, the evolution of each state $\rho _{R}(t)$ can be
written as a Lindblad rate equation \cite{rate,breuer}%
\begin{eqnarray}
\dfrac{d\rho _{R}(t)}{dt}\!\! &=&\!\!\dfrac{-i}{\hbar }[H_{R},\rho
_{R}(t)]-\gamma _{R}(\{D,\rho _{R}(t)\}_{+}-\mathcal{J}[\rho _{R}(t)]) 
\notag \\
&&\!\!-\sum\limits_{R^{\prime }}\phi _{R^{\prime }R}\rho
_{R}(t)+\sum\limits_{R^{\prime }}\phi _{RR^{\prime }}\rho _{R^{\prime }}(t)
\label{LindbladRate} \\
&&\!\!-\sum\limits_{R^{\prime }}\gamma _{R^{\prime }R}\{D,\rho
_{R}(t)\}_{+}+\sum\limits_{R^{\prime }}\gamma _{RR^{\prime }}\mathcal{J}%
[\rho _{R^{\prime }}(t)],  \notag
\end{eqnarray}%
where $\{\cdots \}_{+}$ denotes an anticonmutation operation. The
Hamiltonian reads%
\begin{equation}
H_{R}=\frac{\hbar \omega _{R}}{2}\sigma _{z},\ \ \ \ \ \ \ \ \ \ \ \ \omega
_{R}=\omega _{A}+\delta \omega _{A}^{(R)},  \label{SpectralDiffussion}
\end{equation}%
and the remaining system operators are defined by%
\begin{equation}
D=\sigma ^{\dagger }\sigma /2,\ \ \ \ \ \ \ \ \ \ \ \ \ \mathcal{J}[\bullet
]=\sigma \bullet \sigma ^{\dagger }.  \label{DandJ}
\end{equation}%
The rates associated to the reservoir $\mathcal{W}$\ are given by%
\begin{equation*}
\phi _{RR^{\prime }}=\pi g_{\mathcal{W}}(\Delta _{RR^{\prime }})|\chi
_{RR^{\prime }}^{\Delta _{RR^{\prime }}}|^{2}.
\end{equation*}%
Here, $g_{\mathcal{W}}(\omega )$ is the density of states of $\mathcal{W}$
and $\Delta _{RR^{\prime }}\equiv \omega _{R}-\omega _{R^{\prime }}$ are the
transition frequencies of $\mathcal{U}.$ The rates associated to the field $%
\mathcal{F}$ read ($\gamma _{R}=\gamma _{RR}$)%
\begin{equation}
\gamma _{RR^{\prime }}=2\pi \sum_{\lambda }\int \mathbf{dk}g_{\mathcal{F}%
}^{(R^{\prime })}(\mathbf{k})|\kappa _{RR^{\prime }}^{\mathbf{k},\lambda
}|^{2}\delta (|\mathbf{k}|c-\omega _{A}).
\end{equation}%
Here, $g_{\mathcal{F}}^{(R)}(\mathbf{k})$ is the density of states of the
quantized electromagnetic field associated to each configurational state. By
writing%
\begin{equation}
g_{\mathcal{F}}^{(R)}(\mathbf{k})\mathbf{dk}=\frac{\omega ^{2}V\varepsilon
_{R}^{3/2}}{2\pi ^{2}c^{3}}d\omega d\Omega ,
\end{equation}%
where $d\Omega $ is the solid angle differential and the contribution $%
\varepsilon _{R}^{3/2}$ follows from the dispersion relation Eq.~(\ref%
{dispersion}), after a standard integration \cite{carmichaelbook} it follows%
\begin{equation}
\gamma _{R}=\frac{1}{4\pi }\frac{4\omega _{A}^{3}}{3\hbar c^{3}}%
|d_{RR}|^{2}\varepsilon _{R}^{1/2},\ \ \ \ \ \ \ \gamma _{RR^{\prime }}=%
\frac{1}{4\pi }\frac{4\omega _{A}^{3}}{3\hbar c^{3}}|d_{RR^{\prime
}}|^{2}\varepsilon _{R^{\prime }}^{1/2}.  \label{RatesElectro}
\end{equation}%
For simplicity, it was assumed that the dipole vector $\mathbf{d}%
_{RR^{\prime }}$ [defined by Eq.~(\ref{DipoloOperator})] can be written as $%
\mathbf{d}_{RR^{\prime }}=d_{RR^{\prime }}\mathbf{\hat{d}},$ where $%
d_{RR^{\prime }}$ is the modulus of the dipole vectors, and the unit vector $%
\mathbf{\hat{d}}$ does not depend on the coefficient $R$ and $R^{\prime }.$
While this simplification forbid us to analyze the angular dependence of the
scattered radiation, the general case can be easily worked out from the
present treatment.

The previous expressions rely on the condition $\Delta _{RR^{\prime }}\ll
\omega _{A},$ i.e., the transition frequencies of $\mathcal{U}$ are much
smaller than the optical frequency $\omega _{A}.$ To simplify the analysis,
in Eq.~(\ref{LindbladRate}) we have discarded any shift Hamiltonian
contribution induced by the microscopic dynamic. Furthermore, since the
fluorophore $\mathcal{S}$ is an optical transition, we have assumed that the
average number of thermal excitations of the electromagnetic field at the
characteristic frequency $\omega _{A}$ are much smaller than one, i.e., $%
\bar{n}(\omega _{A},T)\ll 1,$ where $\bar{n}(\omega ,T)=[\exp (\hbar \omega
/k_{B}T)-1]^{-1},$ $k_{B}$ being the Boltzmann constant and $T$ the
temperature associated to the electromagnetic field $\mathcal{F}.$ This
inequality allows to discard in Eq.~(\ref{LindbladRate}) the contributions
that leads to thermal excitations in $\mathcal{S}.$

When the fluorophore is subjected to a resonant external laser field of
frequency $\omega _{L},$ the system Hamiltonian becomes $H_{R}\rightarrow
H_{R}+H_{\mathrm{laser}},$ with 
\begin{equation}
H_{\mathrm{laser}}=\frac{\hbar \Omega _{0}}{2}(\sigma ^{\dagger }e^{-i\omega
_{L}t}+\sigma e^{+i\omega _{L}t}).  \label{Laser}
\end{equation}%
As before, the operators $\sigma ^{\dag }$ and $\sigma ^{\dag }$ are the
raising and lowering operators acting on the states $\{\left\vert
a\right\rangle ,\left\vert b\right\rangle \}.$ The system-laser detuning is
given by%
\begin{equation}
\delta =\omega _{L}-\omega _{A}.  \label{detuning}
\end{equation}%
The Rabi frequency $\Omega _{0}$ reads 
\begin{equation}
\Omega _{0}\equiv \sum_{R}\Omega _{R}\left\vert R\right\rangle \left\langle
R\right\vert ,  \label{Rabi_R}
\end{equation}%
where each $\Omega _{R}$ measure the system-laser coupling for each
configurational state. 
\begin{figure}[tb]
\includegraphics[bb=36 428 792 690,angle=0,width=8.7cm]{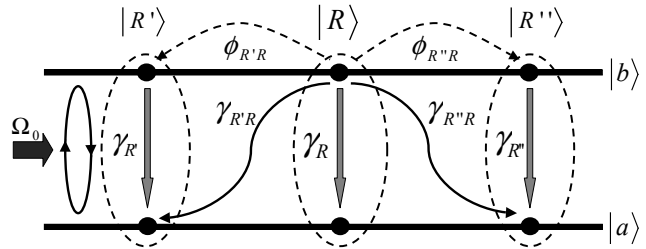}
\caption{Scheme associated to the fluorophore evolution Eq.~(\protect\ref%
{LindbladRate}), see text. }
\label{Figura1_SMS}
\end{figure}

The Lindblad rate equation Eq.~(\ref{LindbladRate}) is the central result of
this section. It effectively describes the action of the nanoscopic
fluctuating environment over the fluorophore. The scheme of Figure 1
symbolically represents all processes associated to this equation. The first
line of Eq.~(\ref{LindbladRate}) describe the self-system dynamic for each
configurational state $\left\vert R\right\rangle .$ The decay rate \{$\gamma
_{R}$\}, the natural transition frequency \{$\omega _{R}$\}, as well as the
Rabi frequency \{$\Omega _{R}$\} may depend on the bath state. The second
line describe transitions between the configurational states (with rates \{$%
\phi _{R^{\prime }R}$\}), whose dynamic does not depend on the state of the
fluorophore (property symbolically represented in Fig.~1 by the surrounding
ellipse). Finally, the third line describes configurational transitions that
are attempted by a transition between the upper and lower states of the
system (rates \{$\gamma _{R^{\prime }R}$\}). In Sec. IV, a detailed analysis
of each contribution and its associated \textquotedblleft kinetic
environment evolution\textquotedblright\ [i.e., the dynamics of $P_{R}(t),$
Eq.~(\ref{ConfigurationalPopulation})] is presented.

As neither the initial conditions or the dynamics [Eq.~(\ref{LindbladRate})]
introduce any coherence between the bath macrostates, their density matrix
can be written as $\rho _{\mathcal{U}}(t)=\sum\nolimits_{R}P_{R}(t)\left%
\vert R\right\rangle \left\langle R\right\vert .$ From the relation $\mathrm{%
Tr}_{\mathcal{U}}[O_{\mathcal{U}}\rho _{\mathcal{U}}(t)]=\mathrm{Tr}_{%
\mathcal{U}}[O_{\mathcal{U}}(t)\rho _{\mathcal{U}}(0)],$ it follows that the
dynamics of operators\ $O_{\mathcal{U}}(t)$\ acting on $\mathcal{H}_{%
\mathcal{U}}$ is classical and dictated by the probabilities $P_{R}(t).$ On
the other hand, the system dynamic arise after \textquotedblleft tracing
out\textquotedblright\ [Eq.~(\ref{RhoS})] all internal transitions between
the configurational states of the reservoir (see Fig.~1). Therefore, the
evolution of the system density matrix must be highly non-Markovian. In
fact, taking into account the results presented in Ref.~\cite{rate}, it
follows%
\begin{equation}
\frac{d\rho _{\mathcal{S}}(t)}{dt}=\mathcal{L}_{H}\rho _{\mathcal{S}%
}(t)+\int_{0}^{t}d\tau \mathbb{L}(t-\tau )\rho _{\mathcal{S}}(\tau ).
\label{master}
\end{equation}%
$\mathcal{L}_{H}$ defines the system unitary dynamic. The equation that
defines the superoperator $\mathbb{L}$ can be explicitly written in a
Laplace domain in terms of the propagator associated to Eq.~(\ref%
{LindbladRate}) (see Eqs.~(58) and (62) in Ref.~\cite{rate}).

\section{Scattered field observables}

In SMS experiments, the fluorophore dynamic is indirectly probed by
subjecting the system to laser radiation [Eq.~(\ref{Laser})] and measuring
the scattered electromagnetic field. Therefore, while the master Eq.~(\ref%
{LindbladRate}) completely characterizes the system dynamic, one is mainly
interested in observables associated to the scattered laser field. In
general, these observables can be written as a function of the electric
field. For example, the flux of energy $S_{\mathcal{F}}$ per unit area per
unit time (module of the Pointyng vector) reads \cite{landau}%
\begin{equation}
S_{\mathcal{F}}(t)=\frac{1}{2}\frac{c}{4\pi }|\mathbf{E}(t)\times \mathbf{H}%
(t)^{\dag }|=\frac{c}{8\pi }\sqrt{\frac{\varepsilon }{\mu }}|\mathbf{E}%
(t)|^{2},  \label{pointing}
\end{equation}%
where the second equality follows from the relation between $\mathbf{E}(t)$
and $\mathbf{H}(t)$ [see Eq.~(\ref{Campos})].

The time evolution of the electric field\ operator follows from a Heisenberg
evolution with respect to the total Hamiltonian Eq.~(\ref{Htotal}), i.e., $%
(d/dt)\mathbf{E}(t)=(i/\hbar )[H_{\mathcal{T}},\mathbf{E}(t)].$ Taking into
account Eq.~(\ref{Electric}), the time dependence of $\mathbf{E}(t)$ can be
obtained from the evolution of the operators $a_{\mathbf{k}\lambda
}^{(R)}\equiv a_{\mathbf{k}\lambda }^{(R)}\left\vert R\right\rangle
\left\langle R\right\vert .$ We get%
\begin{equation}
\frac{da_{\mathbf{k}\lambda }^{(R)}(t)}{dt}\simeq -i\omega _{\mathbf{k}%
}^{(R)}a_{\mathbf{k}\lambda }^{(R)}(t)-i\sum_{R^{\prime }}\kappa
_{RR^{\prime }}^{\mathbf{k},\lambda ^{\ast }}\sigma _{R^{\prime }R}(t),
\label{akevolution}
\end{equation}%
where $\omega _{\mathbf{k}}^{(R)}$ and the coefficients $\kappa _{RR^{\prime
}}^{\mathbf{k},\lambda }$ are defined by Eqs.~(\ref{dispersion}) and (\ref%
{Acoples}) respectively. Furthermore, the $\mathcal{SU}$ operator $\sigma
_{R^{\prime }R}$ is defined by%
\begin{equation}
\sigma _{R^{\prime }R}\equiv \sigma \left\vert R^{\prime }\right\rangle
\left\langle R\right\vert .
\end{equation}%
Eq.~(\ref{akevolution}) relies on a set of approximations consistent with
the effective representation of the reservoir. Since the dielectric constant
of each configurational state is well defined, any non-diagonal non-linear
coupling between the field modes is discarded. Terms arising from $H_{%
\mathcal{U}}^{\prime }$ [Eq.~(\ref{U_Hamiltonian})] are also disregarded
because they only introduce a small modification to the natural (optical)
frequency of each mode.

The dynamics of $a_{\mathbf{k}\lambda }^{(R)}(t)$ can be written as the
addition of two contributions, each one associated to the homogeneous and
inhomogeneous terms in Eq.~(\ref{akevolution}). Then, the electric field is
written as \cite{carmichaelbook,breuerbook}%
\begin{equation}
\mathbf{E}(t)=\mathbf{E}_{\mathrm{f}}(t)+\mathbf{E}_{\mathrm{s}}(t).
\end{equation}%
The contribution $\mathbf{E}_{\mathrm{f}}(t)$ defines the free evolution of
the field. In terms of positive and negative frequency contributions, $%
\mathbf{E}_{\mathrm{f}}(t)=\mathbf{E}_{\mathrm{f}}^{(+)}(t)+\mathbf{E}_{%
\mathrm{f}}^{(-)}(t),$ it is defined by%
\begin{equation}
\mathbf{E}_{\mathrm{f}}^{(+)}(t)=i\sum_{R,\mathbf{k},\lambda }\left\vert
R\right\rangle \left\langle R\right\vert \sqrt{\frac{\hbar \omega _{\mathbf{k%
}}^{(R)}}{2V\varepsilon _{R}}}\hat{e}_{\mathbf{k}\lambda }e^{-i(t\omega _{%
\mathbf{k}}^{(R)}-\mathbf{k\cdot r})}a_{\mathbf{k}\lambda }(0).
\end{equation}%
The scattered field contribution $\mathbf{E}_{\mathrm{s}}(t)$, associated to
the inhomogeneous term in Eq.~(\ref{akevolution}), after some algebra \cite%
{carmichaelbook} can be written as $\mathbf{E}_{\mathrm{s}}(t)=\mathbf{E}_{%
\mathrm{s}}^{(+)}(t)+\mathbf{E}_{\mathrm{s}}^{(-)}(t),$ where%
\begin{equation}
\mathbf{E}_{\mathrm{s}}^{(+)}(t)=\frac{\omega _{A}^{2}}{4\pi c^{2}r}[(%
\mathbf{\hat{d}}\times \mathbf{\hat{r}})\times \mathbf{\hat{r}]}%
\sum_{RR^{\prime }}d_{RR^{\prime }}\sigma _{R^{\prime }R}(t-r/c_{R}),
\label{ElectricScattered}
\end{equation}%
with $\mathbf{\hat{r}\equiv r}/|\mathbf{r}|,$ $r\equiv |\mathbf{r}|,$ and $%
c_{R}=c/\sqrt{\varepsilon _{R}}.$ As before [Eq.~(\ref{RatesElectro})], for
simplicity we assumed that $\mathbf{d}_{R^{\prime }R}=d_{R^{\prime }R}%
\mathbf{\hat{d}}.$

The previous expression allows to obtain the electric field in terms of
operators defined in the system and configurational Hilbert spaces.
Consistently with the effective representation of the reservoir, it is
rewritten as%
\begin{equation}
\mathbf{E}_{\mathrm{s}}^{(+)}(t)=\sum_{RR^{\prime }}\mathbf{E}_{\mathrm{s}%
(R|R^{\prime })}^{(+)(t)}[\left\vert R^{\prime }\right\rangle \left\langle
R\right\vert ](t),  \label{ElectricConclusion}
\end{equation}%
where each contribution reads%
\begin{equation}
\mathbf{E}_{\mathrm{s}(R|R^{\prime })}^{(+)(t)}=\frac{\omega _{A}^{2}}{4\pi
c^{2}r}[(\mathbf{\hat{d}}\times \mathbf{\hat{r}})\times \mathbf{\hat{r}]}\
d_{RR^{\prime }}\sigma _{R^{\prime }}^{\mathrm{c}}(t-r/c_{R}).
\label{ElectricConditional}
\end{equation}%
Since the background electromagnetic field [Eq.~(\ref{Electric})] does not
involve any coherence between the configurational states, the operator $%
[\left\vert R^{\prime }\right\rangle \left\langle R\right\vert ](t)$
appearing in Eq.~(\ref{ElectricConclusion}) must be read as follows. It
labels all contributions to $\mathbf{E}_{\mathrm{s}}^{(+)}(t)$ that at time $%
t$ are attempted by the configurational transition $\left\vert R^{\prime
}\right\rangle \rightarrow \left\vert R\right\rangle ,$ ($R\neq R^{\prime }$%
). Then, the \textquotedblleft conditional operator\textquotedblright\ $%
\mathbf{E}_{\mathrm{s}(R|R^{\prime })}^{(+)(t)}$ defines the electric field
restricted to the condition that at time $t$ the reservoir is in the
configurational state $\left\vert R^{\prime }\right\rangle $ and change to
the state $\left\vert R\right\rangle .$ Similarly, the diagonal
contributions $\mathbf{E}_{\mathrm{s}(R|R)}^{(+)(t)}$ define the electric
field \textquotedblleft \textit{given}\textquotedblright\ that at time $t$
the reservoir is in the configurational state $\left\vert R\right\rangle .$
Consistently, the conditional operator $\sigma _{R}^{\mathrm{c}}(t)$ gives
the evolution of the system operator $\sigma (t)$ under the same condition.

Any scattered field observable must be written in terms of the conditional
system operators $\mathbf{E}_{\mathrm{s}(R|R^{\prime })}^{(+)(t)}.$ Below,
we characterize the field correlations.

\subsection{Correlations}

When the scattered field is measured with photoelectric detectors, the usual
observables can be written in terms of two time (normal order) correlations 
\cite{breuerbook,carmichaelbook,milburn,scully,loudon}%
\begin{equation}
C_{1}(\tau )\equiv \lim_{t\rightarrow \infty }:\overline{\mathbf{E}_{\mathrm{%
s}}^{(-)}(t)\mathbf{E}_{\mathrm{s}}^{(+)}(t+\tau )}:,  \label{C1Electrico}
\end{equation}%
as well as%
\begin{equation}
C_{2}(\tau )\equiv \lim_{t\rightarrow \infty }:\overline{\mathbf{E}_{\mathrm{%
s}}^{(-)}(t)\mathbf{E}_{\mathrm{s}}^{(-)}(t+\tau )\mathbf{E}_{\mathrm{s}%
}^{(+)}(t+\tau )\mathbf{E}_{\mathrm{s}}^{(+)}(t)}:.  \label{C2Electrico}
\end{equation}%
The overbar denotes quantum average with respect to the total initial
density matrix. The symbol $:(\cdots ):$ takes into account the right
interpretation of Eq.~(\ref{ElectricScattered}) and denotes a summation over
all internal configurational paths defined through the conditional
contributions $\mathbf{E}_{\mathrm{s}(R|R^{\prime })}^{(+)(t)},$ Eq.~(\ref%
{ElectricConditional}). Then, the correlations are explicitly written as%
\begin{equation}
C_{1}(\tau )=\lim_{t\rightarrow \infty }\sum_{R_{1},R_{2}}\sum_{R_{3},R_{4}}%
\overline{\mathbf{\tilde{E}}_{\mathrm{s}(R_{2}|R_{1})}^{(-)(t)}\mathbf{%
\tilde{E}}_{\mathrm{s}(R_{4}|R_{3})}^{(+)(t+\tau )}},
\label{C1ElectricoOrdernada}
\end{equation}%
and%
\begin{equation}
C_{2}(\tau )=\lim_{t\rightarrow \infty }\!\!\sum_{\substack{ R_{1},R_{2}  \\ %
R_{3},R_{4}}}\!\overline{\mathbf{\tilde{E}}_{\mathrm{s}
(R_{2}|R_{1})}^{(-)(t)}\mathbf{\tilde{E}}_{\mathrm{s}(R_{4}|R_{3})}^{(-)(t+
\tau )}\mathbf{\tilde{E}}_{\mathrm{s}(R_{4}|R_{3})}^{(+)(t+\tau )}\mathbf{\ 
\tilde{E}}_{\mathrm{s}(R_{2}|R_{1})}^{(+)(t)}},  \label{C2ElectricoOrdernada}
\end{equation}%
where $\mathbf{\tilde{E}}_{\mathrm{s}(R|R^{\prime })}^{(\pm)(t)}\equiv 
\mathbf{\ E}_{\mathrm{s}(R|R^{\prime })}^{(\pm)(t)}\varepsilon _{R^{\prime
}}^{1/4}.$ This definition guarantees that both correlations can be related
to observables defined in terms of energy (photon) fluxes [see Eq.~(\ref%
{pointing})]. It also takes into account that before the transition $%
\left\vert R^{\prime }\right\rangle \rightarrow \left\vert R\right\rangle ,$
the dielectric constant of the bath is $\varepsilon _{R^{\prime }}.$

By using Eq.~(\ref{ElectricConditional}) and the rate expressions Eq.~(\ref%
{RatesElectro}), the first order correlation can be written as%
\begin{equation}
C_{1}(\tau )=f(\mathbf{\hat{r}})\frac{\hbar \omega _{A}}{4\pi cr^{2}}%
\sum_{R,R^{\prime }}(\tilde{\gamma}_{R}\tilde{\gamma}_{R^{\prime
}})^{1/2}\lim_{t\rightarrow \infty }\overline{\sigma _{R}^{\dag \mathrm{c}%
}(t)\sigma _{R^{\prime }}^{\mathrm{c}}(t+\tau )},  \label{C1TauDesarrollada}
\end{equation}%
where for brevity we define $f(\mathbf{\hat{r}})\equiv \lbrack (\mathbf{\hat{%
d}}\times \mathbf{\hat{r}})\times \mathbf{\hat{r}}]^{2},$ and%
\begin{equation}
\tilde{\gamma}_{R}\equiv \gamma _{R}+\sum_{R^{\prime }}\gamma _{R^{\prime
}R}.  \label{RateNormalizada}
\end{equation}%
Eq.~(\ref{C1TauDesarrollada}) can be considered as a natural generalization
of the expression corresponding to the Markovian case \cite{Markov}.
Similarly, the correlation of the (conditional) raising and lowering
operators can also be obtained from the system density matrix evolution
[Eq.~(\ref{LindbladRate})] after invoking to a quantum regression theorem.
For Lindblad rate equations it reads \cite{QRT}%
\begin{equation}
\overline{O_{1}(t)A(t+\tau )O_{2}(t)}\!=\!\sum_{RR^{\prime }}\mathrm{Tr}_{%
\mathcal{S}}\{A(e^{\tau \hat{\mathcal{L}}})_{RR^{\prime }}[O_{2}\rho
_{R^{\prime }}(t)O_{1}]\},  \label{QRT}
\end{equation}%
where $O_{1},$ $O_{2}$ and $A$ are arbitrary system operators. Each
contribution [indexed by $R,R^{\prime }],$ defines a conditional average of
the involved operators: at time $t,$ the configurational bath state is $%
\left\vert R^{\prime }\right\rangle ,$\ while at time $(t+\tau ),$ it is $%
\left\vert R\right\rangle .$ $\hat{\mathcal{L}}$ denotes the generator of
the Lindblad rate equation, 
\begin{equation}
\rho _{R}(t)=\sum_{R^{\prime }}(e^{t\hat{\mathcal{L}}})_{RR^{\prime }}\rho
_{R^{\prime }}(0).  \label{LindbladRateGenerator}
\end{equation}%
From Eqs.~(\ref{QRT}) and Eq.~(\ref{C1TauDesarrollada}) it follows%
\begin{equation}
C_{1}(\tau )=f(\mathbf{\hat{r}})\frac{\hbar \omega _{A}}{4\pi cr^{2}}%
\sum_{RR^{\prime }}\sqrt{\tilde{\gamma}_{R}\tilde{\gamma}_{R^{\prime }}}%
\mathrm{Tr}_{\mathcal{S}}\{\sigma (e^{\tau \hat{\mathcal{L}}})_{RR^{\prime
}}[\rho _{R^{\prime }}^{\infty }\sigma ^{\dagger }]\},  \label{C1Rate}
\end{equation}%
where $\rho _{R}^{\infty }\equiv \lim_{t\rightarrow \infty }\rho _{R}(t).$
By using the same calculations steps, the second order correlation reads%
\begin{eqnarray}
C_{2}(\tau )\! &=&\!\Big[f(\mathbf{\hat{r}})\frac{\hbar \omega _{A}}{4\pi
cr^{2}}\Big]^{2}\!\sum_{RR^{\prime }}\!\tilde{\gamma}_{R}\mathrm{Tr}_{%
\mathcal{S}}\Big\{\sigma ^{\dagger }\sigma (e^{\tau \hat{\mathcal{L}}%
})_{RR^{\prime }}  \label{C2Rate} \\
&&\times \Big[\gamma _{R^{\prime }}\sigma \rho _{R^{\prime }}^{\infty
}\sigma ^{\dag }+\tsum_{R^{\prime \prime }}\gamma _{R^{\prime }R^{\prime
\prime }}\sigma \rho _{R^{\prime \prime }}^{\infty }\sigma ^{\dag }\Big]%
\Big\}.  \notag
\end{eqnarray}%
The expressions Eqs.~(\ref{C1Rate}) and (\ref{C2Rate}) are the central
results of this section. They allow to characterize observables such as the
spectrum of the radiated field as well as the intensity-intensity
correlation.

\subsection{Spectrum}

The spectral intensity radiation (in units of energy $\hbar \omega _{A}$)
per unit of solid angle \cite{carmichaelbook,breuerbook} is defined by the
dimensionless expression%
\begin{equation}
S(\omega )=\frac{1}{2\pi }\int_{-\infty }^{+\infty }d\tau e^{i(\omega
-\omega _{L})\tau }\Big(\frac{\hbar \omega _{A}}{4\pi cr^{2}}\Big)%
^{-1}C_{1}(\tau ),  \label{Espectro}
\end{equation}%
where $\omega _{L}$ is the frequency of the laser excitation, [Eq.~(\ref%
{Laser})]. As usually $\lim_{\tau \rightarrow \infty }C_{1}(\tau )\neq 0,$
the spectrum can be split in a coherent and incoherent contributions 
\begin{equation}
S(\omega )=f(\mathbf{\hat{r}})[S_{coh}(\omega )+(2\pi )^{-1}S_{inc}(\omega
)].  \label{EspectroEspliteado}
\end{equation}%
$S_{coh}(\omega )$ consists in a Dirac delta term 
\begin{equation}
S_{coh}(\omega )\equiv S_{coh}^{(0)}\delta (\omega _{L}),
\label{DiracCoherente}
\end{equation}%
that measures the scattered radiation emitted at the frequency of the laser
excitation. From Eq.~(\ref{C1Rate}), it follows%
\begin{equation}
S_{coh}^{(0)}=\sum_{RR^{\prime }}\sqrt{\tilde{\gamma}_{R}\tilde{\gamma}%
_{R^{\prime }}}\left\langle b\right\vert \rho _{R}^{\infty }\left\vert
a\right\rangle \left\langle a\right\vert \rho _{R^{\prime }}^{\infty
}\left\vert b\right\rangle .  \label{SCoherente}
\end{equation}%
The incoherent contribution can be written as%
\begin{equation}
S_{inc}(\omega )=\Big[\tilde{C}_{1}(u)|_{-i(\omega -\omega _{L})}+\tilde{C}%
_{1}(u)|_{i(\omega -\omega _{L})}\Big],  \label{SIncoherente}
\end{equation}%
where $\tilde{C}_{1}(u)=\int_{0}^{\infty }d\tau e^{-u\tau }\tilde{C}%
_{1}(\tau )$ is the Laplace transform of the function $\tilde{C}_{1}(\tau
)\equiv \lbrack \hbar \omega _{A}f(\mathbf{\hat{r}})/4\pi
cr^{2}]^{-1}C_{1}(\tau ).$

In a Markovian limit, i.e., when the configurational space is
unidimensional, from Eq.~(\ref{SIncoherente}) it is possible to write the
spectrum as an addition of three Lorentzian functions (Mollow triplet) whose
widths and heights depend on the natural decay of the system and the Rabi
frequency \cite{breuerbook,carmichaelbook,milburn,scully}. In the general
non-Markovian case, the spectrum also has a strong dependence on the
parameters that define the environment fluctuations (see next sections). In
spite of these dissimilarities, in both cases the spectrum is mainly related
to the dynamic behavior of the system coherences [see Eq.~(\ref{SCoherente}%
)].

\subsection{Intensity-Intensity correlation}

The normalized intensity-intensity correlation \cite{milburn} reads%
\begin{equation}
g_{2}(\tau )=\lim_{t\rightarrow \infty }\frac{:\overline{\mathbf{I}(t+\tau )%
\mathbf{I}(t)}:}{\overline{\mathbf{I}(t)}^{2}}=\frac{C_{2}(\tau )}{%
|C_{1}(0)|^{2}},
\end{equation}%
where $\mathbf{I}(t)=\mathbf{E}_{\mathrm{s}}^{(-)}(t)\mathbf{E}_{\mathrm{s}%
}^{(+)}(t)\mathbf{\ }$is the intensity operator. From Eqs.~(\ref{C1Rate})
and (\ref{C2Rate}) it follows%
\begin{equation}
g_{2}(\tau )=\frac{1}{(\mathrm{I}_{\mathrm{st}})^{2}}\sum_{RR^{\prime }}%
\tilde{\gamma}_{R}\left\langle b\right\vert (e^{\tau \hat{\mathcal{L}}%
})_{RR^{\prime }}[a_{R^{\prime }}^{\infty }]\left\vert b\right\rangle ,
\label{GTWO}
\end{equation}%
where 
\begin{equation}
a_{R^{\prime }}^{\infty }=\Big\{\gamma _{R^{\prime }}\left\langle
b\right\vert \!\rho _{R^{\prime }}^{\infty }\!\left\vert b\right\rangle
+\sum_{R^{\prime \prime }}\gamma _{R^{\prime }R^{\prime \prime
}}\left\langle b\right\vert \!\rho _{R^{\prime \prime }}^{\infty
}\!\left\vert b\right\rangle \Big\}\!\left\vert a\right\rangle
\!\left\langle a\right\vert ,
\end{equation}%
and the normalization constant reads 
\begin{equation}
\mathrm{I}_{\mathrm{st}}=\sum\limits_{R}\tilde{\gamma}_{R}\left\langle
b\right\vert \rho _{R}^{\infty }\left\vert b\right\rangle .
\label{IntensidadEstacionaria}
\end{equation}%
As in the Markovian case \cite{carmichaelbook},\ Eq.~(\ref{GTWO})
corresponds to the probability density of detecting one photon\ in the
stationary regime ($\lim_{t\rightarrow \infty }$)\ and a second one in the
interval $(\tau ,\tau +d\tau ).$ The factor $a_{R^{\prime }}^{\infty }$
takes into account all possible emission paths that leave the system in the
ground state and the reservoir in the configurational state $\left\vert
R^{\prime }\right\rangle .$ The sum over the index $R$ takes into account
all photon emissions that happen in the interval $(\tau ,\tau +d\tau )$ and
leave the reservoir in the state $\left\vert R\right\rangle .$ The
normalization factor $\mathrm{I}_{\mathrm{st}}$\ defines the average
stationary intensity emitted by the fluorophore.

\section{Photon counting statistics: a generating function approach}

In most of the SMS experiments the scattered laser radiation is measured
with photon detectors. Then, the photon counting statistics is also an usual
observable. As in standard fluorescent systems, the probability $P_{n}(t)$
of detecting $n$ photons up to time $t$ follows from a Mandel formula \cite%
{carmichaelbook,breuerbook}. Here, it is generalized as%
\begin{equation}
P_{n}(t)=\frac{1}{n!}:\overline{\Big[\int_{0}^{t}dt^{\prime }\mathbf{\hat{I}}%
(t^{\prime })\Big]^{n}\exp \Big[-\int_{0}^{t}dt^{\prime }\mathbf{\hat{I}}%
(t^{\prime })\Big]}:,  \label{Mandel}
\end{equation}%
where the normalized intensity operator $\mathbf{\hat{I}}(t)=(\hbar \omega
_{A}/4\pi cr_{d}^{2})^{-1}\mathbf{E}_{\mathrm{s}}^{(-)}(t)\mathbf{E}_{%
\mathrm{s}}^{(+)}(t)$ has units of photon flux. $r_{d}$ denotes the distance
between the fluorophore and the detector. $:(\cdots ):$ denotes both an
usual (normal) time ordering and a summation over all internal
configurational paths, whose definition follows straightforwardly from Eqs.~(%
\ref{C1ElectricoOrdernada}) and (\ref{C2ElectricoOrdernada}).

While Eq.~(\ref{Mandel}) allows to characterize the probabilities $P_{n}(t),$
a simpler technique to calculate these objects is provided by a generating
function approach \cite{cook,mukamelPRA}. This very well known technique was
also used in the context of SMS when dealing with stochastic Bloch equations 
\cite{zheng,FLHBrown,he,BarkaiExactMandel} and related approximations. In
contrast, here we formulate the generating function approach \cite%
{cook,mukamelPRA} on the basis of one of the central results of this
contribution, i.e., Eq.~(\ref{LindbladRate}). Added to its broad generality,
the present formulation avoid the use of any stochastic calculus.

By writing the system density matrix as 
\begin{equation}
\rho _{\mathcal{S}}(t)=\sum\nolimits_{n=0}^{\infty }\rho ^{(n)}(t),
\end{equation}%
where each state $\rho ^{(n)}(t)$ corresponds to the system state
conditioned to $n-$photon detection events \cite{plenio,carmichaelbook}, the
probability of counting $n$-photons up to time $t$ reads 
\begin{equation}
P_{n}(t)=\mathrm{Tr}_{\mathcal{S}}[\rho ^{(n)}(t)].  \label{Pn(t)}
\end{equation}%
A \textquotedblleft generating operator\textquotedblright\ \cite{cook} is
defined by%
\begin{equation}
G(t,s)\equiv \sum\nolimits_{n=0}^{\infty }s^{n}\rho ^{(n)}(t),
\label{Gdefinicion}
\end{equation}%
where $s$ is an extra real parameter. This operator also encodes the system
dynamic, $\rho _{\mathcal{S}}(t)=G(t,s)|_{s=1}.$ The conditional states $%
\rho ^{(n)}(t)$ can be decomposed into the contributions associated to each
configurational state of the reservoir, leading to the expression%
\begin{equation}
G(t,s)=\sum\nolimits_{n=0}^{\infty }s^{n}\sum\nolimits_{R}\rho
_{R}^{(n)}(t)\equiv \sum\nolimits_{R}G_{R}(t,s).  \label{GeneratorOperator}
\end{equation}%
Each matrix $\rho _{R}^{(n)}(t)$ defines the state of the system under the
condition that, at time $t,$ $n-$photon detection events happened and the
configurational state of the environment is $\left\vert R\right\rangle .$
Consistently, each contribution $G_{R}(t,s)$ defines the (conditional)
generating operator \textquotedblleft \textit{given}\textquotedblright\ that
the reservoir is in the configurational state $\left\vert R\right\rangle .$
Its evolution, from Eq.~(\ref{LindbladRate}), can be written as 
\begin{eqnarray}
\dfrac{dG_{R}(t,s)}{dt}\! &=&\!\dfrac{-i}{\hbar }[H_{R},G_{R}(t,s)]-\tilde{%
\gamma}_{R}\{D,G_{R}(t,s)\}_{+}  \notag \\
&&\!+\sum\limits_{R^{\prime }}\phi _{RR^{\prime }}G_{R^{\prime
}}(t,s)\!-\!\sum\limits_{R^{\prime }}\phi _{R^{\prime }R}G_{R}(t,s)
\label{Gr_Evolution} \\
&&\!+s\gamma _{R}\mathcal{J}[G_{R}(t,s)]+s\sum\limits_{R^{\prime }}\gamma
_{RR^{\prime }}\mathcal{J}[G_{R^{\prime }}(t,s)],  \notag
\end{eqnarray}%
where $\tilde{\gamma}_{R}$ is defined by Eq.~(\ref{RateNormalizada}). Notice
that the parameter $s$ is introduced in all terms (third line) associated to
a photon detection event, i.e., those proportional to $\sigma \bullet \sigma
^{\dagger }$ \cite{plenio,carmichaelbook}.

In the context of SMS \cite{zheng,FLHBrown,he,BarkaiExactMandel}, the matrix
elements of $G(t,s),$ in an interaction representation with respect to $%
\hbar \omega _{L}\sigma _{z},$ are usually denoted as 
\begin{subequations}
\label{MatrixElements}
\begin{eqnarray}
\mathcal{U}(t,s) &\equiv &\frac{1}{2}(G_{ab}e^{-i\omega
_{L}t}+G_{ba}e^{i\omega _{L}t}), \\
\mathcal{V}(t,s) &\equiv &\frac{1}{2i}(G_{ab}e^{-i\omega
_{L}t}-G_{ba}e^{i\omega _{L}t}), \\
\mathcal{W}(t,s) &\equiv &\frac{1}{2}(G_{bb}-G_{aa}), \\
\mathcal{Y}(t,s) &\equiv &\frac{1}{2}(G_{bb}+G_{aa}),  \label{YGeneratriz}
\end{eqnarray}%
[$G_{pq}\equiv \left\langle p\right\vert G(t,s)\left\vert q\right\rangle ],$
while their evolution is called \textquotedblleft generalized optical Bloch
equation.\textquotedblright\ From Eq.~(\ref{GeneratorOperator}), it is
possible to write each matrix element as a sum over the parameter $R.$ In
Appendix A, we provide the evolution of each component [Eq.~(\ref%
{OpticalBloch})].

After getting the matrix elements of the generating operator, Eq.~(\ref%
{MatrixElements}), the photon counting process can be characterized in a
standard way \cite{vanKampen}. From the definition of the generating
operator, Eq.~(\ref{Gdefinicion}), the photon counting probabilities Eq.~(%
\ref{Pn(t)}) follows as 
\end{subequations}
\begin{equation}
P_{n}(t)=\frac{2}{n!}\left. \frac{\partial ^{n}}{\partial s^{n}}\mathcal{Y}%
(t,s)\right\vert _{s=0}.  \label{Pn}
\end{equation}%
The function $\mathcal{Y}(t,s)$ also allows to calculate all factorial
moments%
\begin{equation}
\bar{N}^{(k)}(t)\equiv \sum_{n=0}^{\infty }n(n-1)\cdots (n-k+1)P_{n}(t),
\label{factorial}
\end{equation}%
in terms of its derivatives%
\begin{equation}
\bar{N}^{(k)}(t)=2\left. \frac{\partial ^{k}}{\partial s^{k}}\mathcal{Y}%
(t,s)\right\vert _{s=1}.
\end{equation}%
Furthermore, the first two moments of the photon counting process, $%
\overline{N^{k}}(t)\equiv \sum_{n=0}^{\infty }n^{k}P_{n}(t),$ ($k=1,2$),
read 
\begin{subequations}
\label{numero}
\begin{eqnarray}
\overline{N}(t) &=&2\left. \frac{\partial }{\partial s}\mathcal{Y}%
(t,s)\right\vert _{s=1},  \label{NumeroMedio} \\
\overline{N^{2}}(t) &=&2\left. \frac{\partial ^{2}}{\partial s^{2}}\mathcal{Y%
}(t,s)\right\vert _{s=1}+2\left. \frac{\partial }{\partial s}\mathcal{Y}%
(t,s)\right\vert _{s=1}.\ \ \ \ 
\end{eqnarray}%
Both moments encode important information about the scattered radiation. The
line shape of the fluorophore system is defined by \cite{barkaiChem} 
\end{subequations}
\begin{equation}
\mathrm{I}(\omega _{L})\equiv \lim_{t\rightarrow \infty }\frac{d}{dt}%
\overline{N}(t),  \label{espectro}
\end{equation}%
while the Mandel factor is defined as 
\begin{equation}
Q(t)\equiv \frac{\overline{N^{2}}(t)-\overline{N}^{2}(t)}{\overline{N}(t)}-1.
\label{MandelDefinition}
\end{equation}%
As is well known \cite{breuerbook,carmichaelbook,milburn,scully}, it allows
to determining the sub- or super-Poissonian character of the photon counting
process. In Appendix B we show the consistency between the generating
function approach, the Mandel formula Eq.~(\ref{Mandel}), and the results
obtained in the previous section. In Appendix C, it is shown that the
stationary Mandel factor 
\begin{equation}
Q_{st}\equiv \lim_{t\rightarrow \infty }Q(t),  \label{MandelEstacionario}
\end{equation}%
can be obtained in an exact analytical way after solving the evolution Eq.~(%
\ref{Gr_Evolution}) in the Laplace domain.

\section{Examples}

In this section, different processes covered by Eq.~(\ref{LindbladRate}) are
analyzed. The examples are classified in accordance with the evolution of
the environment fluctuations, i.e., the evolution of the configurational
populations $P_{R}(t),$ Eq.~(\ref{ConfigurationalPopulation}). In each case,
observables such as the spectrum Eq.~(\ref{Espectro}), intensity-intensity
correlation, Eq.~(\ref{GTWO}), line shape, Eq.~(\ref{espectro}), and Mandel
factor, Eq.~(\ref{MandelDefinition}), can be calculated in an exact
analytical way. In fact, in a Laplace domain, both the evolution of the
auxiliary density states, Eq.~(\ref{LindbladRate}), and the evolution of the
auxiliary generating operators, Eq.~(\ref{Gr_Evolution}), become algebraic
linear equations.

\subsection{Self environment fluctuations}

When the nano-reservoir only has dense manifold of states, the fluctuations
between the configurational states must be governed by a classical master
equation \cite{vanKampen}. This case is covered by taking 
\begin{eqnarray}
\dfrac{d\rho _{R}(t)}{dt} &=&\frac{-i}{\hbar }[H_{R},\rho _{R}(t)]-\gamma
_{R}(\{D,\rho _{R}(t)\}_{+}-\mathcal{J}[\rho _{R}(t)])  \notag \\
&&-\sum\limits_{R^{\prime }}\phi _{R^{\prime }R}\rho
_{R}(t)+\sum\limits_{R^{\prime }}\phi _{RR^{\prime }}\rho _{R^{\prime }}(t),
\label{LindbladSelf}
\end{eqnarray}%
which arises from Eq.~(\ref{LindbladRate}) under the condition $\mathbf{d}%
_{RR^{\prime }}=0,$ $R\neq R^{\prime },$ implying the vanishing of the rates 
$\{\gamma _{RR^{\prime }}\}$\ [Eq.~(\ref{RatesElectro})]. From Eq.~(\ref%
{LindbladSelf}), the evolution of the configurational populations, Eq.~(\ref%
{ConfigurationalPopulation}), reads 
\begin{equation}
\frac{d}{dt}P_{R}(t)=-\sum\limits_{R^{\prime }}\phi _{R^{\prime
}R}P_{R}(t)+\sum\limits_{R^{\prime }}\phi _{RR^{\prime }}P_{R^{\prime }}(t).
\label{Classical}
\end{equation}%
In the context of SMS, this evolution defines the \textquotedblleft kinetic
dynamic\textquotedblright\ of the environment. Consistently, this equation,
and in consequence the transitions between the configurational states $%
\{\left\vert R\right\rangle \}_{R=1}^{R_{\max }}$, does not depend on the
state of the system. Notice that in the effective microscopic
representation, the configurational transitions are induced by the reservoir 
$\mathcal{W},$ Eq.~(\ref{U_Hamiltonian}).

While the fluctuations between the configurational states are defined by the
classical evolution Eq.~(\ref{Classical}), the quantum dynamic of the
system, for each bath state, is defined by the diagonal contribution in Eq.~(%
\ref{LindbladSelf}). Depending on its properties, different processes are
recovered.

\subsubsection{Spectral fluctuations}

A nano-environment consisting on molecules at very low temperatures may
induce random fluctuations in the natural frequency of a fluorophore \cite%
{barkaiChem}. This situation can be described by the present formalism by
assuming that the natural frequency of the system, Eq.~(\ref%
{SpectralDiffussion}), is the unique parameter that depends on the
configurational bath states. Consequently, each configurational bath state
has associated a different system transition frequency, $\omega _{R}=\omega
_{A}+\delta \omega _{R}.$ Equivalently, each configurational state
\textquotedblleft induces\textquotedblright\ a spectral shift defined by the
parameter $\delta \omega _{R}.$ The population $P_{R}(t)$ gives the
probability that, at time $t,$ the natural frequency of the system is $%
\omega _{R}.$ The coefficients associated to the Rabi frequency, Eq.~(\ref%
{Rabi_R}), must be chosen all the same. Similarly, the decay rates
associated to each bath state, Eq.~(\ref{RatesElectro}), are also all the
same. Therefore, the system is characterized by one single Rabi frequency
and a natural decay rate, i.e.,%
\begin{equation}
\Omega _{R}=\Omega ,\ \ \ \ \ \ \ \ \ \ \ \ \ \ \ \gamma _{R}=\gamma .
\end{equation}

In the following analysis, the configurational space is assumed two
dimensional, $R=1,2.$ Furthermore, the\ problem is restricted to a
symmetrical one. The characteristic parameters read 
\begin{subequations}
\begin{eqnarray}
\delta \omega _{A}^{(1)} &=&+\delta \omega ,\ \ \ \ \ \ \ \ \ \ \ \ \ \ \ \
\ \ \phi _{21}=\phi , \\
\delta \omega _{A}^{(2)} &=&-\delta \omega ,\ \ \ \ \ \ \ \ \ \ \ \ \ \ \ \
\ \ \phi _{12}=\phi .
\end{eqnarray}%
With these assumptions, from the results of Appendix C, it is possible to
characterize both the line shape, Eq.~(\ref{espectro}), and the stationary
Mandel factor, Eq.~(\ref{MandelEstacionario}), in an exact analytical way.
The final expressions recover the results obtained by He and Barkai in Ref.~%
\cite{BarkaiExactMandel}, where the natural frequency of the fluorophore is
modeled through a stochastic two-state process. 
\begin{figure}[tb]
\includegraphics[bb=5 5 844 1300,angle=0,width=7 cm]{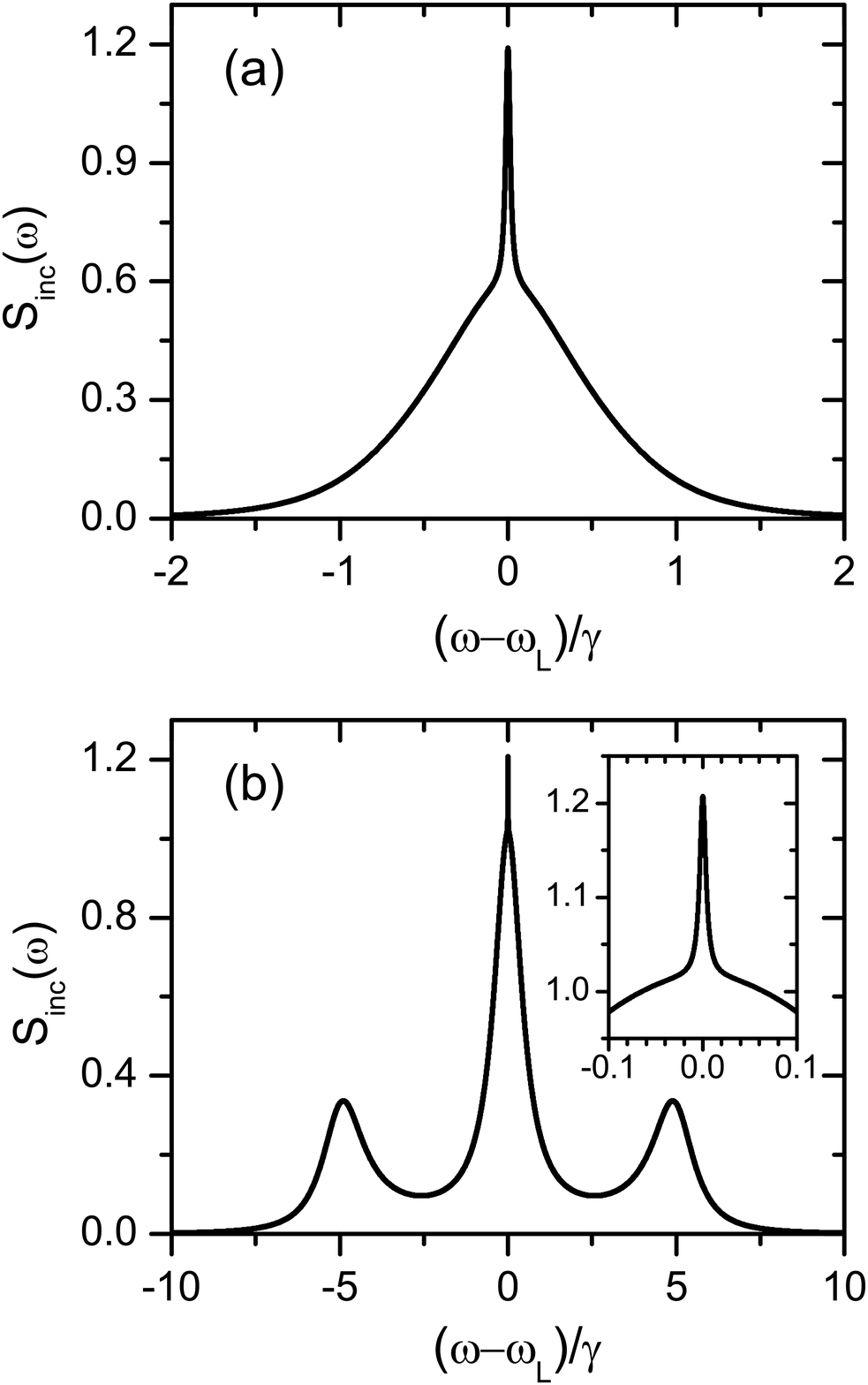}
\caption{Optical spectra [Eq.~(\protect\ref{SIncoherente})] for a
fluorophore system subject to spectral fluctuations in the regime $\protect%
\phi \ll \protect\delta \protect\omega \ll \protect\gamma \simeq \Omega .$
(a) The parameters, in units of the decay rate $\protect\gamma ,$ are $%
\Omega =\protect\gamma /\protect\sqrt{2},$ $\protect\delta \protect\omega =%
\protect\gamma /10,$ and $\protect\phi =\protect\gamma /125.$ (b) The
parameters are $\Omega =5\protect\gamma ,$ $\protect\delta \protect\omega =%
\protect\gamma /10,$ and $\protect\phi =\protect\gamma /500.$ The inset
shows the narrow spectral peak. In both cases, the laser frequency is in
resonance with the system, $\protect\omega _{L}=\protect\omega _{A}.$}
\end{figure}

Now, going beyond the possibilities of any previous approach, in the
following figures we characterize the spectrum of the scattered radiation
[Eq.~(\ref{SIncoherente})]. Its exact analytical expression is not provided
due to its extension. In all cases, the system-laser detuning [Eq.~(\ref%
{detuning})] is zero, i.e., $\omega _{L}=\omega _{A}.$ When the
characteristic time between configurational transitions, $\phi ^{-1},$ is
the small time scale of the problem, the spectrum shape strongly depends on
the relation between the natural decay $\gamma ,$ the Rabi frequency $\Omega
,$\ and the spectral shift $\delta \omega .$ In Fig.~2, the condition $\phi
\ll \delta \omega \ll \gamma \simeq \Omega $ is satisfied. Then, the
spectral fluctuations induced by the bath slightly modify quantum dynamic of
the system. Nevertheless, the spectrum shape presents strong deviations with
respect to the Markovian case.

In Fig.~2a, as $\Omega <\gamma ,$ the spectrum only consists in one
Lorentzian component (Rayleigh peak) \cite{plenio}. In contrast to standard
(Markovian) fluorescent systems, here a central narrow peak is clearly
visible. It must not be confused with the coherent Dirac delta contribution
that appears in both Markovian and non-Markovian systems [Eq.~(\ref%
{DiracCoherente})]. In fact, as in three level systems \cite{plenio}, its
origin can be related to the dynamics of the system coherences \cite{despuez}%
. The spectrum develops a narrow peak whenever the dynamic of the coherences
slowly fluctuates between two different regimes. The width of the narrow
peak is given by the addition of the constant rates that define the
fluctuations between the different dynamical regimes \cite{plenio}. Here, as 
$\phi \ll \delta \omega ,$ each dynamical regime is defined by the spectral
shifts $\pm \delta \omega $ associated to each configurational bath state.
The \textquotedblleft blinking\textquotedblright\ between these two regimes
is governed by the rate $\phi .$ Therefore, the width of the narrow peak is $%
2\phi ,$ which allows us to read a central property of the environment
fluctuations from the optical spectra.%
\begin{figure}[tb]
\includegraphics[bb=7 5 844 1302,angle=0,width=7 cm]{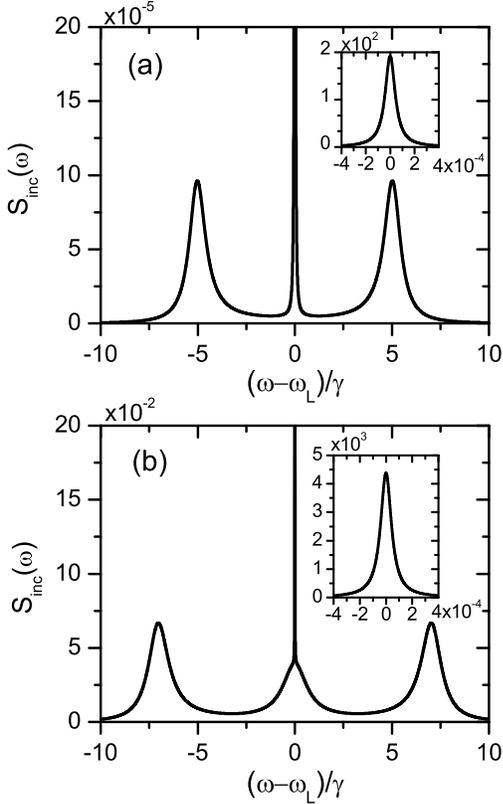}
\caption{Optical spectra [Eq.~(\protect\ref{SIncoherente})] for a
fluorophore system subject to spectral fluctuations in the regimes $\protect%
\phi \ll \protect\gamma \simeq \Omega \ll \protect\delta \protect\omega $
(a) and $\protect\phi \ll \protect\gamma \ll \Omega \simeq \protect\delta 
\protect\omega $ (b). In (a) the parameters, in units of the decay rate $%
\protect\gamma ,$ are $\Omega =\protect\gamma /\protect\sqrt{2},$ $\protect%
\delta \protect\omega =5\protect\gamma ,$ and $\protect\phi =\protect\gamma %
/(4\times 10^{4}).$ In (b) the parameters are $\Omega =5\protect\gamma ,$ $%
\protect\delta \protect\omega =5\protect\gamma ,$ and $\protect\phi =\protect%
\gamma /(4\times 10^{4}).$ In both cases, the laser frequency is in
resonance with the system, $\protect\omega _{L}=\protect\omega _{A}.$ The
insets corresponds to the central narrow peaks.}
\end{figure}

In Markovian fluorescent systems, when $\Omega >\gamma $ the laser strongly
couples the population and coherences of the system, leading to oscillations
that induce two extra peaks in the optical spectra (Mollow triplet) \cite%
{carmichaelbook}. Here, this phenomenon is shown in Fig.~2b, where the
parameters satisfy $\phi \ll \delta \omega \ll \gamma <\Omega .$ As the Rabi
frequency $\Omega $ is much larger than the spectral shifts $\delta \omega ,$
the peaks appear at $\pm \Omega .$ Consistently with the parameter values,
the extra narrow peak (inset) is also associated with the coherences
blinking-like evolution.

In Fig.~3a, the condition $\phi \ll \gamma \simeq \Omega \ll \delta \omega $
is satisfied. Therefore, the quantum system dynamic is mainly governed by
the spectral shifts induced by the environment. Consistently, the spectrum
develops two peaks at $\pm $ $\delta \omega .$ In contrast with the Mollow
triplet, which is associated to the Rabi oscillations, here the central
Rayleigh peak ($\omega _{L}=\omega _{A}$) is almost absent. It only appears
a narrow central peak (inset) related to the coherence dynamics.

By maintaining the value of all parameters, in Fig.~3b, the laser intensity
was increased such that $\phi \ll \gamma \ll \Omega \simeq \delta \omega .$
In this situation, there is a competition between the Rabi oscillations and
the spectral shifts induced by the laser and the reservoir respectively.
Notice that the extra peaks do not appear at $\pm \Omega $ [Fig.~2b] neither
at $\pm \delta \omega $ [Fig.~3a]. Furthermore, in contrast to standard
fluorescent systems, the height of the central Rayleigh peak is smaller than
the height of the lateral peaks. The inset shows the narrow central peak.
Since the parameters of the environmental fluctuations are the same than in
Fig.~3a, the central narrow peaks have the same widths, i.e., $2\phi .$ 
\begin{figure}[tb]
\includegraphics[bb=5 5 840 1263,angle=0,width=7 cm]{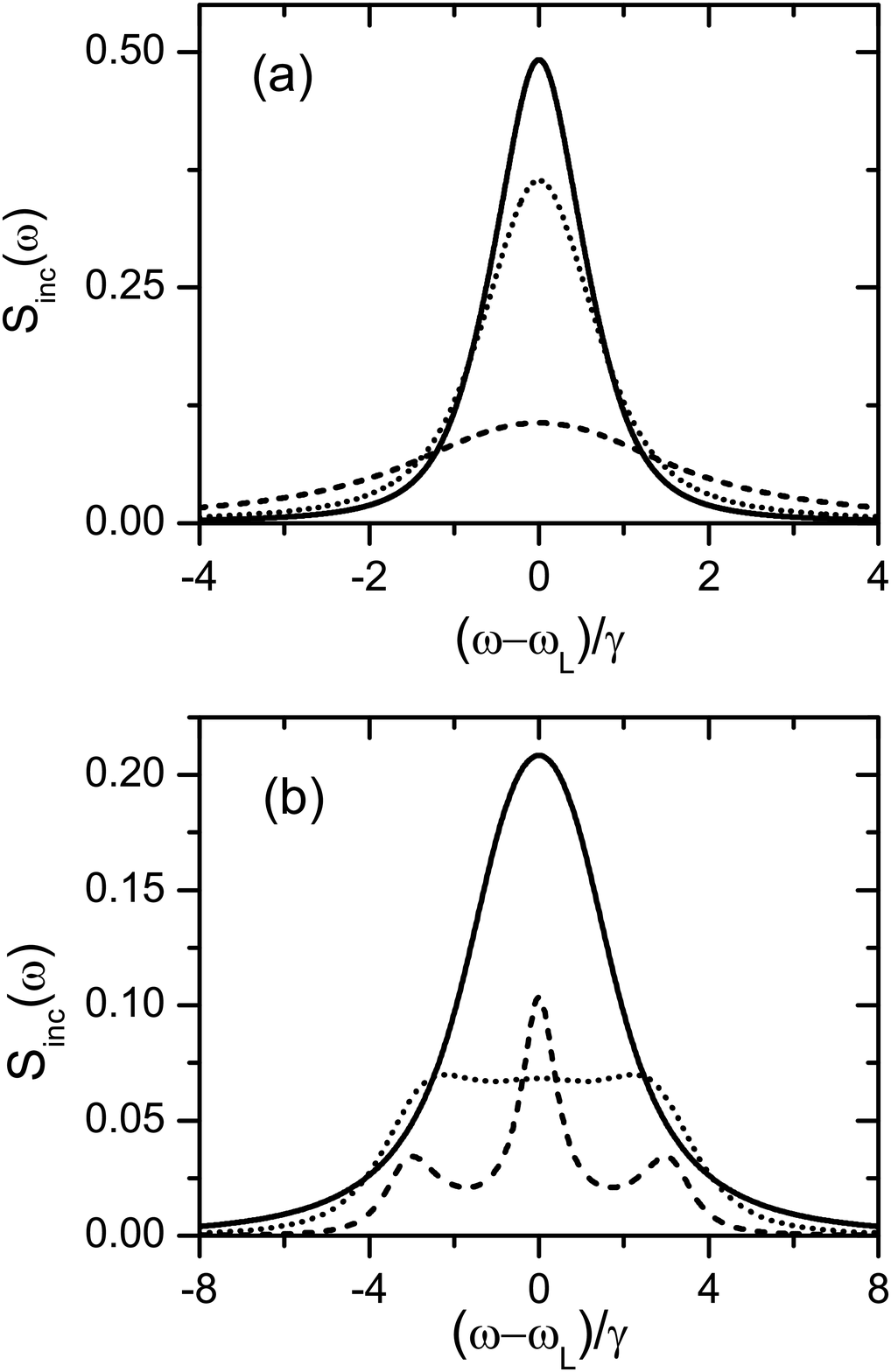}
\caption{Optical spectra [Eq.~(\protect\ref{SIncoherente})] for a
fluorophore system subject to spectral fluctuations. In (a) the parameters,
in units of the decay rate $\protect\gamma ,$ are $\Omega =\protect\gamma /%
\protect\sqrt{2},$ $\protect\delta \protect\omega =5\protect\gamma ,$ and $%
\protect\phi =10\protect\gamma $ (dashed line), $\protect\phi =50\protect%
\gamma $ (dotted line), $\protect\phi =125\protect\gamma $ (solid line). In
(b) the parameters are $\Omega =\protect\gamma ,$ $\protect\delta \protect%
\omega =3\protect\gamma ,$ and $\protect\phi =0.25\protect\gamma $ (dashed
line), $\protect\phi =\protect\gamma $ (dotted line), $\protect\phi =4%
\protect\gamma $ (solid line). In both cases, the laser frequency is in
resonance with the system, $\protect\omega _{L}=\protect\omega _{A}.$ }
\end{figure}

A broad kind of spectrum behavior may arise when the environment fluctuation
time $\phi ^{-1}$ is not the small time scale of the problem. In this
regime, the coherences dynamic loose its dichotomic character and
consequently the spectrum does not develop any central narrow peak \cite%
{despuez}. In Fig.~4a the reservoir fluctuations are faster than the system
dynamic, $\gamma \simeq \Omega \simeq \delta \omega \ll \phi .$ The spectrum
develops the well-known effect of motional narrowing, i.e., by increasing
the rate $\phi $ the optical spectrum becomes narrow. In Fig.~4b the rate $%
\phi $ is increased around the intermediate regime $\gamma \simeq \phi .$ As
the parameters of the system and the environment fluctuations are of the
same order, a small variation of $\phi $ lead to strong changes in the
spectral shape. This property can also be found in the photon counting
statistics (see Fig.~5 in Ref.~\cite{BarkaiExactMandel}).

\subsubsection{Lifetime fluctuations}

Chemical or physical conformational changes of the supporting
nano-environment may modify the lifetime of the fluorophore \cite%
{barkaiChem,michel,jung}. This situation is covered by taking $(\gamma
_{RR^{\prime }}=0)$ 
\end{subequations}
\begin{equation}
\Omega _{R}=\Omega ,\ \ \ \ \ \ \ \ \ \ \ \ \ \ \ \delta \omega _{R}=0.
\label{Life}
\end{equation}%
Then, each configurational state $\left\vert R\right\rangle $ has associated
the natural decay $\gamma _{R}.$ The classical evolution Eq.~(\ref{Classical}%
) describe the transitions, with rates $\phi _{R^{\prime }R},$ between the
different decay rates felt by the fluorophore.

\subsubsection{Molecules diffusing in solution}

The present formalism may also be applied for describing the statistical
properties of radiation patterns produced by molecules diffusing in a
solution \cite{michel,Elson}. If the intensity of the laser field varies
appreciably along the diffusing space, the Rabi frequency (which depends
linearly on the external laser power) must be considered dependent on
position. Then, the configurational space can be associated to the physical
space where the diffusion process happens. This situation is covered by
taking $(\gamma _{RR^{\prime }}=0)$ 
\begin{equation}
\gamma _{R}=\gamma ,\ \ \ \ \ \ \ \ \ \ \ \ \ \ \ \ \delta \omega _{R}=0.
\end{equation}
Each coefficient $\Omega _{R}$ [Eq.~(\ref{Rabi_R})] measures the laser
intensity at each position labeled by the index $R.$ The evolution Eq.~(\ref%
{Classical}) describes the molecule diffusion. For an homogeneous coupling
between first neighbors, $\phi _{RR\pm 1}=\phi ,$ a standard diffusion
process with diffusion coefficient $D_{0}=\Delta x^{2}\phi ,$ is recovered. $%
\Delta x$ is the discretization step in physical space.

\subsection{Light assisted environment fluctuations}

When the local nano-reservoir surrounding the system only consists in a few
degrees of freedom, its Hilbert space is defined by a discrete (finite) set
of states. Therefore, each one may represents a different configurational
state. While there exist different Lindblad rate equations that may model
this situation (see end of this section), here we consider the case 
\begin{eqnarray}
\dfrac{d\rho _{R}(t)}{dt}\!\! &=&\!\!\dfrac{-i}{\hbar }[H_{R},\rho
_{R}(t)]\!-\!\gamma _{R}(\{D,\rho _{R}(t)\}\!_{+}-\!\mathcal{J}[\rho
_{R}(t)])  \notag \\
&&\!\!\!\!\!-\!\sum\limits_{R^{\prime }}\!\gamma _{R^{\prime }\!R}\{\!D,\rho
_{R}(t)\}\!_{+}+\!\sum\limits_{R^{\prime }}\!\gamma _{RR^{\prime }}\mathcal{%
\ J}\![\rho _{R^{\prime }}(t)].  \label{LightAssisted}
\end{eqnarray}%
This evolution follows from Eq.~(\ref{LindbladRate}) after taking $\{\phi
_{RR^{\prime }}\}=0,$ condition consistent with the finite structure of the
reservoir, i.e., its dynamic is unable to induces self (internal)
fluctuations \cite{vanKampen}. In the effective microscopic representation,
it follows from the interaction Eq.~(\ref{Htotal}) by uncoupling the
configurational space from the reservoir $\mathcal{W},$ i.e., $H_{\mathcal{UW%
}}=0$ [Eq.~(\ref{Huw})].


Taking into account the interaction Hamiltonian Eq.~(\ref{HSUB}), we deduce
that transitions between the different configurational bath states may only
happen when they are attempted by a photon emission process, i.e., a
transition between the upper and lower system states (see Fig.~1).
Therefore, in contrast with the previous case, here the dynamic of the
configurational populations [Eq.~(\ref{ConfigurationalPopulation})] strongly
depends on the state of the system. From Eq.~(\ref{LightAssisted}) we get 
\begin{equation}
\frac{d}{dt}P_{R}(t)=-\sum\limits_{R^{\prime }}\gamma _{R^{\prime
}R}P_{R}^{(b)}(t)+\sum\limits_{R^{\prime }}\gamma _{RR^{\prime
}}P_{R^{\prime }}^{(b)}(t),  \label{PPMas}
\end{equation}%
where $P_{R}^{(b)}(t)\equiv \left\langle b\right\vert \rho _{R}(t)\left\vert
b\right\rangle .$ The evolution of $P_{R}^{(b)}(t)$ also follows from Eq.~(%
\ref{LightAssisted}), which in turn involves the remaining matrix elements
of all auxiliary states. Therefore, in general it is not possible to write a
simple equation for the evolution of the configurational populations.
Nevertheless, in the limits $\{\gamma _{RR^{\prime }}\}\ll \{\gamma _{R}\}$
and $\{\gamma _{RR^{\prime }}\}\gg \{\gamma _{R}\},$ an evolution similar to
Eq.~(\ref{Classical}) can be obtained. We assume that 
\begin{equation}
\Omega _{R}=\Omega ,\ \ \ \ \ \ \ \ \ \ \ \ \ \ \ \delta \omega _{R}=0,
\end{equation}%
implying that the fluorophore, independently of the configurational state,
always feels the same laser intensity $\Omega ,$ and maintains its
transition frequency $\omega _{A}.$ These conditions allow to model a class
of light assisted processes consistent with different experimental
situations \cite{luzAssisted}.

Under the condition $\{\gamma _{RR^{\prime }}\}\ll \{\gamma _{R}\},$ from
Eq.~(\ref{LightAssisted}) it is simple to realize that before happening a
configurational transition, the fluorophore may emit a large number of
photons. Thus, \textquotedblleft \textit{while}\textquotedblright\ the bath
remains in the configurational state $\left\vert R\right\rangle ,$ the
fluorophore can be well approximated by a Markovian fluorescent system with
decay rate $\tilde{\gamma}_{R}$ [Eq.~(\ref{RateNormalizada})], the radiation
pattern being characterized by the average intensity 
\begin{equation}
I_{R}=\frac{\tilde{\gamma}_{R}\Omega ^{2}}{\tilde{\gamma}_{R}^{2}+2\Omega
^{2}+4\delta ^{2}},  \label{Intensidad_R}
\end{equation}%
where $\delta =\omega _{L}-\omega _{A}$ is the system-laser detuning, Eq.~(%
\ref{detuning}). Consequently, it is possible to approximate 
\begin{equation}
\tilde{\gamma}_{R}P_{R}^{(b)}(t)\simeq I_{R}P_{R}(t).
\label{AproximationLigth}
\end{equation}%
The first term represents the photon flux produced by the system while the
bath remains in the configurational state $\left\vert R\right\rangle .$ This
quantity is approximated by\ the intensity $I_{R}$ multiplied by the
probability of being in the configurational state $\left\vert R\right\rangle
,$ i.e., $P_{R}(t).$ Similarly, a straightforward interpretation of this
approximation follows by noticing that $I_{R}/\tilde{\gamma}_{R}$ gives the
stationary upper population of a Markovian optical transition with
characteristic decay rate $\tilde{\gamma}_{R}.$ Then, Eq.~(\ref%
{AproximationLigth}) implies that, \textquotedblleft \textit{given}%
\textquotedblright\ that the reservoir remains in the state $\left\vert
R\right\rangle ,$ the upper population $P_{R}^{(b)}(t)$ quickly reaches the
stationary value $I_{R}/\tilde{\gamma}_{R}.$

By introducing Eq.~(\ref{AproximationLigth}) in the evolution Eq.~(\ref%
{PPMas}), it follows the classical evolution%
\begin{equation}
\frac{d}{dt}P_{R}(t)\simeq -\sum\limits_{R^{\prime }}\Gamma _{R^{\prime
}R}P_{R}(t)+\sum\limits_{R^{\prime }}\Gamma _{RR^{\prime }}P_{R^{\prime
}}(t),  \label{LigthPRBlinking}
\end{equation}%
where the transition rates read%
\begin{equation}
\Gamma _{R^{\prime }R}=\gamma _{R^{\prime }R}\left( \frac{\Omega ^{2}} {%
\tilde{\gamma}_{R}^{2}+2\Omega ^{2}+4\delta ^{2}}\right) .  \label{hophop}
\end{equation}%
These expressions generalize the results presented in Ref.~\cite{luzAssisted}%
. The associated radiation pattern develops a blinking phenomenon, i.e., the
intensity of the scattered radiation randomly changes between the set of
values $\{I_{R}\}$ associated to each configurational state. These changes
are governed by the classical evolution Eq.~(\ref{LigthPRBlinking}). The
dependence of the transition rates $\Gamma _{R^{\prime }R}$ on both the
laser intensity and the system-laser detuning [compare with Eq.~(\ref%
{Classical})], gives the light assisted character of the blinking
phenomenon. In contrast with triplet blinking models \cite{barkaiChem,molski}%
, here all rates that govern the transitions between the configurational
states (intensity states) depend on both, the laser intensity and the system
laser detuning, the first property being consistent with different
experimental results \cite{luzAssisted}.

When $\{\gamma _{RR^{\prime }}\}\gg \{\gamma _{R}\},$ in the intermediate
time between two consecutive configurational transitions, the system is only
able to emits a few number of photons. Therefore, the blinking phenomenon is
lost because it is not possible to associate an intensity regime [with value
Eq.~(\ref{Intensidad_R})] to each configurational state. In spite of this
fact, under the replacement $\tilde{\gamma}_{R}\rightarrow (\sum_{R^{\prime
\prime }}\gamma _{R^{\prime \prime }R})$ the approximation Eq.~(\ref%
{AproximationLigth}) is still valid. Since $\{\gamma _{RR^{\prime }}\}\gg
\{\gamma _{R}\},$ it follows $\tilde{\gamma}_{R}\simeq (\sum_{R^{\prime
\prime }}\gamma _{R^{\prime \prime }R}),$ implying that both Eq.~(\ref%
{LigthPRBlinking}) and (\ref{hophop}) remain valid when $\{\gamma
_{RR^{\prime }}\}\gg \{\gamma _{R}\}.$ 
\begin{figure}[tb]
\includegraphics[bb=5 6 840 1263,angle=0,width=7 cm]{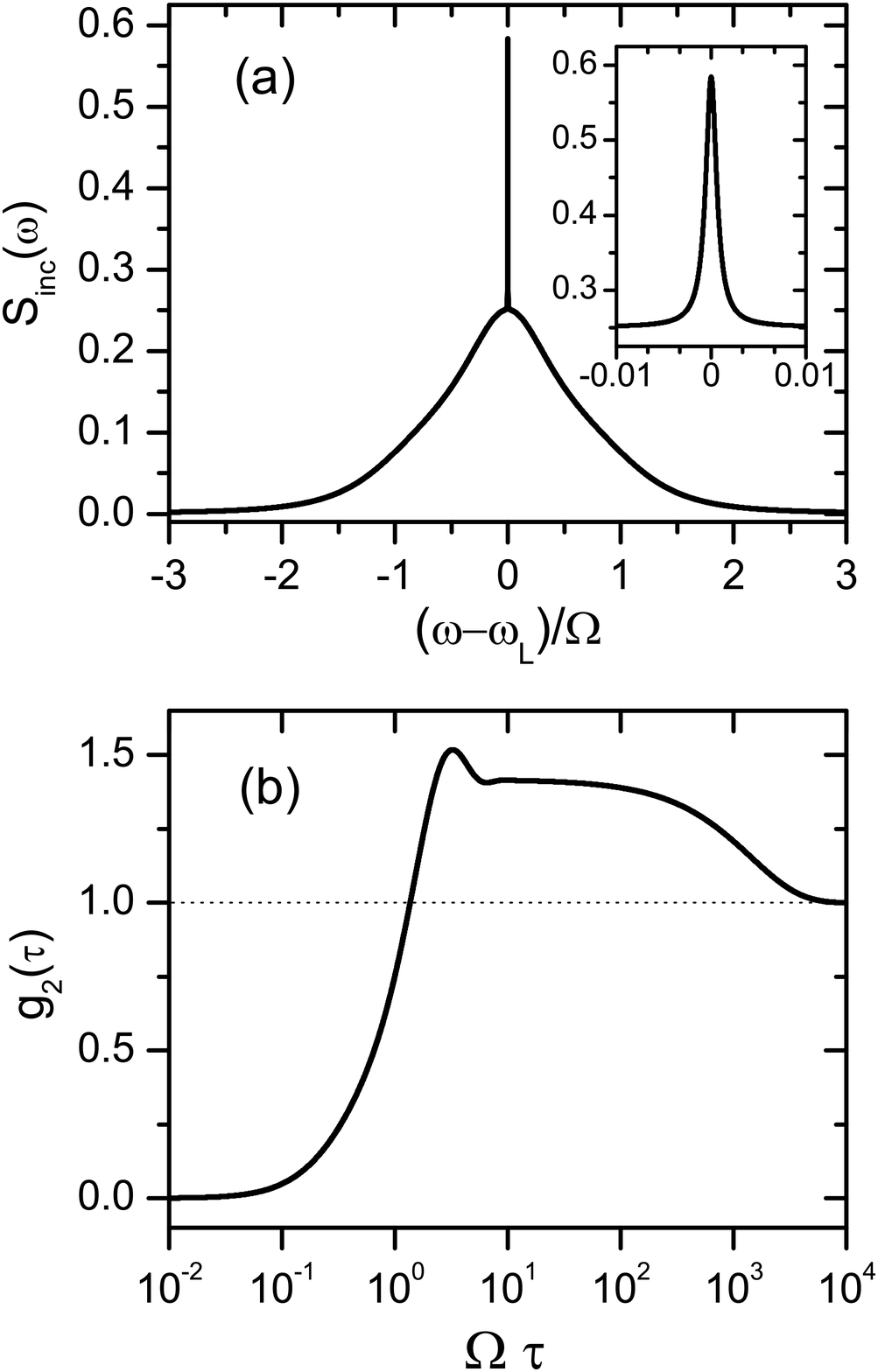}
\caption{Optical spectra [Eq.~(\protect\ref{SIncoherente})] (a) and
intensity-intensity correlation [Eq.~(\protect\ref{GTWO})] (b) for a light
assisted blinking radiation pattern. The configurational space is two
dimensional, $R=1,2$. The parameter values, in units of the Rabi frequency $%
\Omega ,$ are $\protect\gamma _{1}=\Omega ,$ $\protect\gamma _{2}=10\Omega ,$
$\protect\gamma _{21}=0.0015\Omega ,$ $\protect\gamma _{12}=0.02\Omega .$
The laser is in resonance with the system, $\protect\omega _{L}=\protect%
\omega _{A}.$ In (a) the inset shows the central narrow peak.}
\end{figure}

In the following figures, the light assisted blinking phenomena described in
Ref.~\cite{luzAssisted} is characterized through the scattered field
observables. A two dimensional configurational space is assumed ($R=1,2$)
and also the condition $\{\gamma _{RR^{\prime }}\}\ll \{\gamma _{R}\}$ is
satisfied. Fig.~5a shows the optical spectrum [Eq.~(\ref{SIncoherente})].
The Rayleigh central peak is endowed with a narrow peak (inset). As in the
previous analysis, it origin relies on the blinking like behavior of the
system coherences, effect associated to Eq.~(\ref{LigthPRBlinking}). The
width of the narrow peak \cite{despuez}, $\Gamma _{12}+\Gamma _{21},$
depends on both the system-laser detuning and the laser intensity [see Eq.~(%
\ref{hophop})].

Fig.~5b shows the normalized intensity-intensity correlation [Eq.~(\ref{GTWO}%
)]. For short times, this function satisfies $0\leq g_{2}(\tau )<1,$
property associated to photon antibunching. At intermediates times the
correlation satisfies $g_{2}(\tau )>1,$ indicating photon bunching, which in
turn is a direct manifestation of the environment fluctuations. Furthermore, 
$g_{2}(\tau )$ is almost constant during a long period of time. This
property is related with the telegraphic nature (high and low intensity
states) of the scattered radiation field \cite{michel,plenio}.

The central narrow peak in the spectra as well as the photon
bunching-antibunching transition reflected by the intensity-intensity
correlation (Fig.~5) are associated to the blinking property of the
scattered field intensity. Then, it is expected that, independently of the
underlying dynamic, these features also arise whenever the radiated field
fluctuates between two different intensity regimes. For example, the
blinking phenomenon defined by Eq.~(\ref{LigthPRBlinking}) and (\ref{hophop}%
), also can arises from a fluorophore whose environment, independently of
the system state [Eq.~(\ref{Classical})], randomly changes its natural decay
between two different values, i.e., Eq.~(\ref{LindbladSelf}) with the
parameters values Eq.~(\ref{Life}). Consistently, we have checked that under
the mapping 
\begin{subequations}
\label{Mapping}
\begin{eqnarray}
\gamma _{1} &\rightarrow &\tilde{\gamma}_{1}=(\gamma _{1}+\gamma _{21}),\ \
\ \ \ \ \ \ \ \ \phi _{21}=\Gamma _{21}, \\
\gamma _{2} &\rightarrow &\tilde{\gamma}_{2}=(\gamma _{2}+\gamma _{12}),\ \
\ \ \ \ \ \ \ \ \phi _{12}=\Gamma _{12},
\end{eqnarray}%
the spectrum [Eq.~(\ref{SIncoherente})] as well as the intensity correlation
[Eq.~(\ref{GTWO})] associated to Eq.~(\ref{LindbladSelf}) [and Eq.~(\ref%
{Life})] are indistinguishable from those shown in Fig.~5. In spite of these
similarities, both cases lead to very different photon counting processes.%
\begin{figure}[tb]
\includegraphics[bb=3 2 831 600,angle=0,width=8.5 cm]{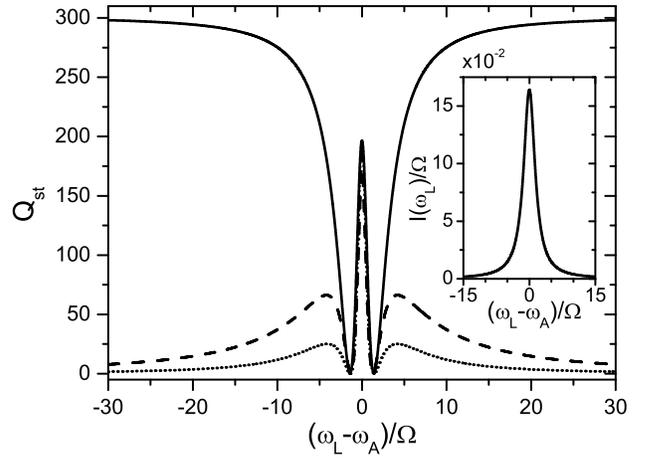}
\caption{Stationary Mandel factor [Eq.~(\protect\ref{MandelEstacionario})]
as a function of the system-laser detuning. For the light assisted process
(full line) the parameters are the same than in Fig. 5. The inset shows the
associated line shape [Eq.~(\protect\ref{espectro})]. For the self
fluctuating environment (dotted line) the parameters are defined from the
mapping Eq.~(\protect\ref{Mapping}). The dashed line corresponds to the
scaling Eq.~(\protect\ref{MapeoTriplete}) with $\protect\delta _{0}=\Omega ,$
$\bar{\Omega}=0.25\Omega ,$ and $\bar{\protect\gamma}_{12}=0.007\Omega .$ }
\end{figure}

In Fig.~6, the photon counting statistics is characterized through the
stationary Mandel factor $Q_{st}$ [see Appendix C] as a function of the
system-laser detuning, i.e., $\delta =\omega _{L}-\omega _{A}$ [Eq.~(\ref%
{detuning})]. While the number of emitted photons (the stationary intensity)
is inversely proportional to $\delta ,$ in the limit of large detuning the
Mandel factor may take arbitrary values. In fact, this object [Eq.~(\ref%
{MandelDefinition})] measures the (normalized) fluctuations in the number of
counted photons around its mean value. For the light assisted dynamic [Eq.~(%
\ref{LightAssisted})] (solid line), the photon statistic becomes
super-Poissonian, i.e., $Q_{st}>0,$ with the asymptotic value
\end{subequations}
\begin{equation}
\lim_{\delta \rightarrow \infty }Q_{st}=\frac{2\gamma _{12}\gamma
_{21}[(\gamma _{1}+\gamma _{21})-(\gamma _{2}+\gamma _{12})]^{2}}{(\gamma
_{12}+\gamma _{21})^{2}(\gamma _{1}\gamma _{12}+\gamma _{2}\gamma
_{21}+2\gamma _{12}\gamma _{21})}>0.  \label{MandelDetuning}
\end{equation}%
Notice that this expression does not depend on the external laser
excitation. This result is expected because the condition $\Omega \ll \delta 
$ is satisfied. Instead, the photon counting statistic for the self
fluctuating environment [Eq.~(\ref{LindbladSelf})] (dotted line) becomes
Poissonian for larger detuning, $\lim_{\delta \rightarrow \infty }Q_{st}=0.$
On the other hand, both dynamics are characterized by the same Mandel factor
when the system-laser detuning is small. The inset shows the line shape Eq.~(%
\ref{espectro}) associated to the light assisted process.

The super-Poissonian character [$Q_{st}>0$] of the photon counting process
in the limit of increasing system-laser detuning has a clear origin. For the
light assisted process, all transition rates $\Gamma _{R^{\prime }R}$ [Eq.~(%
\ref{hophop})] vanishes in the limit $\delta \rightarrow \infty .$ Instead,
for the self-fluctuating environment the transition rates does not vanish in
the same limit. In fact, they are independent of the properties of the
external laser excitation, Eq.~(\ref{Classical}). The vanishing of $%
\lim_{\delta \rightarrow \infty }Q_{st}$ can be recover when the asymptotic
value of one of the transition rates [Eq.~(\ref{hophop})] does not depend on
the system-laser detuning. This property can be achieved by assuming the
following scaling (dashed line in Fig.~6)%
\begin{equation}
\gamma _{12}\rightarrow \gamma _{12}+\bar{\gamma}_{12}(|\delta |/\delta
_{0}),\ \ \ \ \ \ \ \ \ \ \ \ \ \ \ \Omega \rightarrow \Omega +\bar{\Omega}%
(|\delta |/\delta _{0})^{1/2},  \label{MapeoTriplete}
\end{equation}%
where $\delta _{0},$ $\bar{\Omega},$ and $\bar{\gamma}_{12}$ are arbitrary
constants. Both $\gamma _{12}$ and $\Omega $ increase as $\delta $
increases. Consequently, the transition rate $\Gamma _{12}$ does not vanish
by increasing $\delta .$ In fact, $\lim_{\delta \rightarrow \infty }\Gamma
_{12}\simeq \bar{\gamma}_{12}\bar{\Omega}^{2}/(\bar{\gamma}_{12}+4\delta
_{0}^{2}).$ Furthermore, as $\gamma _{2}\ll \gamma _{12},$ \textquotedblleft 
\textit{given}\textquotedblright\ that the bath is in the configurational
state $\left\vert 2\right\rangle ,$ most of the photon emissions are
attempted by the transition $\left\vert 2\right\rangle \rightarrow
\left\vert 1\right\rangle ,$ implying $I_{2}\simeq 0.$ Therefore, with the
scaling Eq.~(\ref{MapeoTriplete}), the dynamic can be mapped with a triplet
blinking modelling \cite{barkaiChem,molski,mandel} where the configurational
state $\left\vert 2\right\rangle $ defines a dark state incoherently coupled
to the system. On the other hand, by assuming valid the condition $\gamma
_{21}\ll \gamma _{1},$ from Eqs.~(\ref{Intensidad_R}) and (\ref{hophop}) it
follows $I_{1}\simeq \gamma _{1}\bar{\Omega}^{2}/(4\delta _{0}|\delta |),$
and $\Gamma _{21}\simeq \gamma _{21}\bar{\Omega}^{2}/(4\delta _{0}|\delta |),
$ which in turn implies that the bath state $\left\vert 1\right\rangle $ can
be associated to the bright intensity states. Finally, from Eq.~(\ref%
{MandelDetuning}), the scaling Eq.~(\ref{MapeoTriplete}) implies $%
\lim_{\delta \rightarrow \infty }Q_{st}\simeq 2\gamma _{21}/\gamma _{1}\ll 1.
$

\subsection{General environment fluctuations}

Having understood the derivation of Eq.~(\ref{LindbladRate}) and their
associated physical processes, one can quickly write down master equations
that introduce more general environment fluctuations. The evolution of the
auxiliary states is written as%
\begin{equation}
\dfrac{d\rho _{R}(t)}{dt}=(\mathcal{L}_{H}^{(R)}+\mathcal{L}_{diag}^{(R)}+%
\mathcal{L}_{cf}^{(R)})[\rho _{R}(t)].
\end{equation}%
The superoperators $\mathcal{L}_{H}^{(R)}$ and $\mathcal{L}_{diag}^{(R)}$
define the unitary and irreversible evolution for each configurational bath
state respectively. Both kind of contributions have been analyzed in the
previous examples. Nevertheless, notice that extra pure dispersive
contributions (phase destroying process) may also be considered. The non
diagonal superoperator $\mathcal{L}_{cf}^{(R)}$introduces the more general
environment fluctuations. It can be written as%
\begin{equation}
\mathcal{L}_{cf}^{(R)}[\rho _{R}]=-\!\sum\limits_{\substack{ R^{\prime }  \\ %
R^{\prime }\neq R}}\frac{\eta _{R^{\prime }R}}{2}\{A^{\dagger }A,\rho
_{R}\}_{+}\!+\!\sum\limits_{\substack{ R^{\prime }  \\ R^{\prime }\neq R}}%
\eta _{RR^{\prime }}A\rho _{R^{\prime }}A^{\dagger },  \label{LGeneral}
\end{equation}%
where the matrix $\eta _{R^{\prime }R}$ defines the characteristic
non-diagonal rates of the problem. $A$ is an arbitrary system operator.

Only when $A=\mathrm{I},$ where the $\mathrm{I}$ is the identity operator,
the environment fluctuations, defined by the evolution of $P_{R}(t),$ do not
depend on the state of the system, recovering Eq.~(\ref{LindbladSelf}). When 
$A=\sigma =\left\vert a\right\rangle \left\langle b\right\vert ,$ the
evolution Eq.~(\ref{LightAssisted}) is recovered, which is associated to
bath fluctuations that are attempted by the system transition $\left\vert
b\right\rangle \rightarrow \left\vert a\right\rangle ,$ i.e., by a photon
emission. When $A=\sigma ^{\dagger }=\left\vert b\right\rangle \left\langle
a\right\vert ,$ the configurational transitions are attempted by a system
transition from the lower to the upper state, $\left\vert a\right\rangle
\rightarrow \left\vert b\right\rangle ,$ i.e., the system absorbs a photon
from the background electromagnetic field. This kind of contributions, which
also arises from the microscopic interaction Eq.~(\ref{HSUB}), were
discarded because at room temperatures the optical thermal excitations are
almost null.

For $A=\left\vert b\right\rangle \left\langle b\right\vert ,$ Eq.~(\ref%
{LGeneral}) describes configurational transitions that can only happen when
the fluorophore is in the upper state. These processes can be
microscopically described by replacing the interaction Hamiltonian $H_{%
\mathcal{UW}},$ Eq.~(\ref{Huw}), by $\left\vert b\right\rangle \left\langle
b\right\vert \otimes H_{\mathcal{UW}}.$ They induce light assisted processes
similar to those described by Eq.~(\ref{LightAssisted}). In particular, when 
$\delta \rightarrow \infty ,$ the stationary photon counting statistics is
also super-Poissonian. Similarly, $A=\left\vert a\right\rangle \left\langle
a\right\vert ,$ describes configurational transitions that only happen when
the system is in the ground state. In both cases, the dynamic induced by
Eq.~(\ref{LGeneral}) introduces a dephasing mechanism that affects the
fluorophore yield.

The formalism does not forbid the situation in which $\eta _{R^{\prime
}R}=0. $ This case was (partially) addressed in Refs. \cite{rapid,mandel},
where the underlying dynamic was defined in terms of a structured reservoir
whose influence can be approximated by a direct sum of Markovian
sub-reservoirs (generalized Born-Markov approximation) \cite{rate,breuer}.
Most of the expressions for the scattered field observables obtained
previously remain valid. Nevertheless, some properties of the radiation
pattern may depend on which way the system-reservoir fluctuations are
measured \cite{despuez}.

\section{Summary and Conclusions}

The main theoretical problem of SMS is to relate the radiation of a single
fluorophore system with the underlying dynamical fluctuations of its local
environment. In this paper, we tackled this problem from a
quantum-electrodynamic treatment based on an open quantum system approach.

The central ingredient of our formalism is the description of the
nano-environment. Instead of considering the microscopic dynamics associated
to each specific situation, after noting that the Hilbert space structure of
the reservoir can not be resolved beyond the experimental resolution, it was
approximated by a set of configurational macrostates, each one taking in
account all microscopic bath configurations that lead to the same system
dynamic. This coarse grained representation allows to define an effective
microscopic description, where the fluorophore and background
electromagnetic field Hamiltonians, as well as their mutual interaction, are
parametrized by the configurational states. From the effective microscopic
Hamiltonian, it was possible to achieve the central goals of this paper.

By tracing out the electromagnetic field and the configurational degrees of
freedom, we derived the Lindblad rate evolution Eq.~(\ref{LindbladRate}),
which encode the statistical behavior of the fluorophore and the reservoir
fluctuations.

In contrast with previous approaches, the scattered electromagnetic field
was characterized from a full quantum-electrodynamic description. The
electric field operator was written in terms of conditional system
operators, Eq.~(\ref{ElectricConditional}). After appealing to a quantum
regression theorem, they allow to expressing the field correlations in terms
of the system density matrix propagator, Eqs.~(\ref{C1Rate}) and (\ref%
{C2Rate}). High order correlations can also be expressed  in a similar way.
These results allow to get simple manageable expressions for observables
like the spectrum, Eq.~(\ref{EspectroEspliteado}), and the
intensity-intensity correlation, Eq.~(\ref{GTWO}).

An extended Mandel formula, Eq.~(\ref{Mandel}), which includes an additional
sum over all configurational transitions, describe the photon counting
statistics. A simpler characterization based on the generating operator Eq.~(%
\ref{GeneratorOperator}) was presented. The line shape and stationary Mandel
factor can be obtained straightforwardly after solving its evolution, Eq.~(%
\ref{Gr_Evolution}), in the Laplace domain.

The examples worked out in the manuscript were classified according to the
properties of the configurational fluctuations. Processes like spectral
diffusion and lifetime fluctuations correspond to bath fluctuations that do
not depend on the system state while light assisted processes are recovered
in the opposite case, i.e., when the environmental fluctuations become
entangled with the system dynamic. From a microscopic point of view, the
former case arises in local environments characterized by dense manifold of
states while the last one arises from reservoirs defined by few degrees of
freedom. A recipe for studying arbitrary bath fluctuations is given by Eq.~(%
\ref{LGeneral}).

The approach allowed to study observables and phenomena that cannot be
easily obtained from formalisms that do not take explicitly into account the
quantum nature of the scattered electromagnetic field. We showed that the
incoherent optical spectra may develops narrow peaks whose width allow to
read the value of some of the characteristic rates that govern the reservoir
fluctuations. A convergence to a super-Poissonian photon counting statistic
(in the limit of large system-laser detuning) in light assisted processes
was analyzed in detail.

The open quantum system approach provides a solid basis for modeling a broad
class of SMS experiments. The effective microscopic dynamic and the
quantum-electrodynamic treatment give the possibility of exploring many open
problems. Since a density matrix description is available, a general
formulation of a quantum jump approach should provide insight into the
properties of the photon-to-photon emission process. The incidence of
non-stationary phenomena and quantum effects in the configurational space
are also interesting situations that can be dealt with the ideas introduced
in this paper.

\section*{Acknowledgments}

The author thank fruitful discussions with Prof. G. Buendia and Prof. V.M.
Kenkre. This work was supported by CONICET, Argentina, as well as from
Consortium of the Americas for Interdisciplinary Science, NSF's
International Division via Grant INT-0336343.

\appendix

\section{Generalized Optical Bloch equation}

The notation of the matrix elements Eq.~(\ref{MatrixElements}) can be
extended trivially to each contribution $G_{R}(t,s)$ in Eq.~(\ref%
{GeneratorOperator}). The evolution of their matrix elements, i.e., the
generalized optical Bloch equation, from Eq.~(\ref{Gr_Evolution}), read 
\begin{widetext}
\begin{subequations}
\label{OpticalBloch}
\begin{eqnarray}
\mathcal{\dot{U}}_{R}(t,s) &=&\delta _{R}\mathcal{V}_{R}(t,s)-\Big[\frac{1}{2%
}\tilde{\gamma}_{R}+\sum\limits_{R^{\prime }}\phi _{R^{\prime }R}\Big]%
\mathcal{U}_{R}(t,s)+\sum_{R^{\prime }}\phi _{RR^{\prime }}\mathcal{U}%
_{R^{\prime }}(t,s), \\
\mathcal{\dot{V}}_{R}(t,s) &=&-\delta _{R}\mathcal{U}_{R}(t,s)-\Omega _{R}%
\mathcal{W}_{R}(t,s)-\Big[\frac{1}{2}\tilde{\gamma}_{R}+\sum_{R^{\prime
}}\phi _{R^{\prime }R}\Big]\mathcal{V}_{R}(t,s)+\sum_{R^{\prime }}\phi
_{RR^{\prime }}\mathcal{V}_{R^{\prime }}(t,s),\ \ \ \ \ \ \ \  \\
\mathcal{\dot{W}}_{R}(t,s) &=&\Omega _{R}\mathcal{V}_{R}(t,s)-\frac{1}{2}(%
\tilde{\gamma}_{R}+s\gamma _{R})\left[ \mathcal{W}_{R}(t,s)+\mathcal{Y}%
_{R}(t,s)\right]  \\
&&-\frac{s}{2}\sum_{R^{\prime }}\gamma _{RR^{\prime }}[\mathcal{W}%
_{R^{\prime }}(t,s)+\mathcal{Y}_{R^{\prime }}(t,s)]-\sum_{R^{\prime }}\phi
_{R^{\prime }R}\mathcal{W}_{R}(t,s)+\sum_{R^{\prime }}\phi _{RR^{\prime }}%
\mathcal{W}_{R^{\prime }}(t,s),  \notag \\
\mathcal{\dot{Y}}_{R}(t,s) &=&-\frac{1}{2}(\tilde{\gamma}_{R}-s\gamma _{R})[%
\mathcal{W}_{R}(t,s)+\mathcal{Y}_{R}(t,s)] \\
&&+\frac{s}{2}\sum\limits_{R^{\prime }}\gamma _{RR^{\prime }}[\mathcal{W}%
_{R^{\prime }}(t,s)+\mathcal{Y}_{R^{\prime }}(t,s)]-\sum_{R^{\prime }}\phi
_{R^{\prime }R}\mathcal{Y}_{R}(t,s)+\sum_{R^{\prime }}\phi _{RR^{\prime }}%
\mathcal{Y}_{R^{\prime }}(t,s),  \notag
\end{eqnarray}
\end{subequations}
\end{widetext}
where $\delta _{R}\equiv (\omega _{L}-\omega _{A})-\delta \omega _{A}^{(R)}$
[Eq.~(\ref{Omega_Cero})], $\Omega _{R}$ and $\tilde{\gamma}_{R}$ being
defined by Eqs.~(\ref{Rabi_R}) and (\ref{RateNormalizada}) respectively.

\section{Intensity-Intensity correlation\ in the generating function approach%
}

In this Appendix, the intensity-intensity correlation Eq.~(\ref{GTWO}) is
obtained from the generating function approach Eq.~(\ref{Gr_Evolution}).
These calculations shows the consistency between the generating function
approach, the Mandel formula Eq.~(\ref{Mandel}), and the results of Sec. III.

The function $\mathcal{Y}(t,s),$ Eq.~(\ref{YGeneratriz}), from the
definition Eq.~(\ref{GeneratorOperator}), can be written as 
\begin{equation}
\mathcal{Y}(t,s)=\frac{1}{2}\mathrm{Tr}_{\mathcal{S}}[G(t,s)]=\frac{1}{2}%
\sum\nolimits_{R}\mathrm{Tr}_{\mathcal{S}}[G_{R}(t,s)].
\end{equation}%
Then, the evolution Eq.~(\ref{Gr_Evolution}) allow us to calculate the time
derivative of the factorial moments $\bar{N}^{(n)}(t),$ Eq.~(\ref{factorial}%
). For $n=1,$ we get%
\begin{equation}
\frac{d}{dt}\bar{N}^{(1)}(t)=\sum\limits_{R}\tilde{\gamma}_{R}\mathrm{Tr}_{%
\mathcal{S}}[\sigma ^{\dag }\sigma \rho _{R}(t)]=\sum\limits_{R}\tilde{\gamma%
}_{R}\left\langle b\right\vert \rho _{R}(t)\left\vert b\right\rangle .
\end{equation}%
Since $\bar{N}^{(1)}(t)=\overline{N}(t)$ [Eq.~(\ref{NumeroMedio})], the line
shape Eq.~(\ref{espectro}) can alternatively be written as%
\begin{equation}
I(\omega _{L})=\sum\limits_{R}\tilde{\gamma}_{R}\left\langle b\right\vert
\rho _{R}^{\infty }\left\vert b\right\rangle ,
\end{equation}%
where $\rho _{R}^{\infty }\equiv \lim_{t\rightarrow \infty }\rho _{R}(t).$
This expression recovers Eq.~(\ref{IntensidadEstacionaria}), showing that
the line shape is proportional to the initial value of the first order
correlation, Eq.~(\ref{C1Electrico}). The same expression follows
straightforwardly from the Mandel formula, Eq.~(\ref{Mandel}). By using the
same procedure, the evolution of the second factorial moment reads%
\begin{equation}
\frac{d}{dt}\bar{N}^{(2)}(t)=2\sum\limits_{R}\tilde{\gamma}_{R}\mathrm{Tr}_{%
\mathcal{S}}\left[ \sigma ^{\dag }\sigma \left. \frac{\partial }{\partial s}%
G_{R}(t,s)\right\vert _{s=1}\right] .  \label{DerivalFactor(2)}
\end{equation}%
The contributions $\left. (\partial /\partial s)G_{R}(t,s)\right\vert
_{s=1}, $ can also be obtained from Eq.~(\ref{Gr_Evolution}),%
\begin{equation}
\frac{\partial }{\partial t}\left. \frac{\partial }{\partial s}%
G_{R}(t,s)\right\vert _{s=1}=\sum\limits_{R^{\prime }}\hat{\mathcal{L}}%
_{RR^{\prime }}\left. \frac{\partial }{\partial s}G_{R^{\prime
}}(t,s)\right\vert _{s=1}+a_{R}(t),  \label{GRS}
\end{equation}%
where $\hat{\mathcal{L}}_{RR^{\prime }}$ is defined by Eq.~(\ref%
{LindbladRateGenerator}) and the inhomogeneous term $a_{R}(t)$ reads%
\begin{equation}
a_{R}(t)=\gamma _{R}\mathcal{J}[\rho _{R}(t)]+\sum\limits_{R^{\prime
}}\gamma _{RR^{\prime }}\mathcal{J}[\rho _{R^{\prime }}(t)].
\end{equation}%
After a formal integration of Eq.~(\ref{GRS}), from Eq.~(\ref%
{DerivalFactor(2)}) $\bar{N}^{(2)}(t)$ can be written as%
\begin{equation}
\bar{N}^{(2)}(t)=2\int_{0}^{t}dt^{\prime }\int_{0}^{t^{\prime }}dt^{\prime
\prime }\digamma (t^{\prime }-t^{\prime \prime },t^{\prime \prime }),
\end{equation}%
where%
\begin{equation}
\digamma (\tau ,t)=\sum\limits_{RR^{\prime }}\tilde{\gamma}_{R}\mathrm{Tr}_{%
\mathcal{S}}\{\sigma ^{\dag }\sigma (e^{\tau \hat{\mathcal{L}}})_{RR^{\prime
}}[a_{R^{\prime }}(t)]\}.  \label{EfE}
\end{equation}%
Notice that this expression, with the exception of the spacial and angular
dependences, recovers Eq.~(\ref{C2Rate}). Now, from the Mandel formula Eq.~(%
\ref{Mandel}), after a standard set of calculations steps \cite%
{loudon,knight,carmichaelbook}, it follows the relation%
\begin{equation}
\bar{N}^{(2)}(t)=2\int_{0}^{t}dt^{\prime }\int_{0}^{t^{\prime }}dt^{\prime
\prime }:\overline{\mathbf{\hat{I}}(t^{\prime }-t^{\prime \prime })\mathbf{%
\hat{I}}(t^{\prime \prime })}:.
\end{equation}%
Therefore, the intensity-intensity correlation reads 
\begin{equation}
g^{(2)}(\tau )=\frac{\lim_{t\rightarrow \infty }\digamma (\tau ,t)}{I(\omega
_{L})^{2}},
\end{equation}%
which from Eq.~(\ref{EfE}) recovers Eq.~(\ref{GTWO}). These results show the
internal consistency of the developed approach.

\section{Stationary Mandel factor}

In this Appendix, the stationary Mandel factor Eq.~(\ref{MandelEstacionario}%
) is obtained from the function $\mathcal{Y}(t,s)$ [Eq.~(\ref{MatrixElements}%
)] in the Laplace domain. From Eq.~(\ref{numero}), the Mandel factor can be
rewritten as%
\begin{equation}
Q(t)=\frac{\mathcal{Y}^{\prime \prime }(t,1)-2[\mathcal{Y}^{\prime
}(t,1)]^{2}}{\mathcal{Y}^{\prime }(t,1)},  \label{MandelYYY}
\end{equation}%
where the accent denotes derivation with respect to $s.$ In the long time
limit, the derivatives of $\mathcal{Y}(t,s)$ behave as 
\begin{subequations}
\label{asimptoticos}
\begin{eqnarray}
\mathcal{Y}^{\prime }(t,1) &\approx &a+bt, \\
\mathcal{Y}^{\prime \prime }(t,1) &\approx &C+At+Bt^{2}.
\end{eqnarray}%
The constant $a$ is zero only when the system begins in its stationary
state. From Eqs.~(\ref{MandelYYY}) and (\ref{asimptoticos}) the stationary
Mandel factor, under the condition $B=2b^{2},$ reads 
\end{subequations}
\begin{equation}
Q_{st}=\frac{A}{b}-4a.
\end{equation}%
The condition $B=2b^{2}$ guarantees the cancellation of terms cuadratic in
time. On the other hand, from the definition Eq.~(\ref{espectro}), the line
shape reads%
\begin{equation}
I(\omega _{L})=2b.  \label{linea}
\end{equation}%
The coefficients associated to the asymptotic time behaviors, Eq.~(\ref%
{asimptoticos}), can be obtained in the Laplace domain. In the limit of $%
u\rightarrow 0,$ it is always possible to write%
\begin{eqnarray}
\mathcal{Y}^{\prime }(u,1) &\approx &\frac{p+qu+O[u^{2}]}{%
Pu^{2}+Qu^{3}+O[u^{4}]}, \\
\mathcal{Y}^{\prime \prime }(u,1) &\approx &\frac{\tilde{p}+\tilde{q}%
u+O[u^{2}]}{\tilde{P}u^{3}+\tilde{Q}u^{4}+O[u^{5}]}.
\end{eqnarray}%
implying the relations%
\begin{eqnarray}
b &=&\frac{p}{P},\ \ \ \ \ \ \ \ \ \ \ \ \ \ \ \ \ \ \ a=\frac{q}{P}-\frac{pQ%
}{P^{2}}, \\
2B &=&\frac{\tilde{p}}{\tilde{P}},\ \ \ \ \ \ \ \ \ \ \ \ \ \ \ \ \ \ A=%
\frac{\tilde{q}}{\tilde{P}}-\frac{\tilde{p}\tilde{Q}}{\tilde{P}^{2}}.
\end{eqnarray}%
Then, the line shape and stationary Mandel factor can be obtained in an
exact way after solving the\ generalized optical Bloch equation [Eq.~(\ref%
{OpticalBloch})] in the Laplace domain.


\begin{thebibliography}{99}
\bibitem{Markov} In the Markovian case \cite{carmichaelbook,breuerbook}, the
correlations are $C_{1}(\tau )=f(\mathbf{\hat{r}})(\hbar \omega _{A}/4\pi
cr^{2})\gamma \lim_{t\rightarrow \infty }\overline{\sigma ^{\dagger
}(t)\sigma (t+\tau )},$ while the second order correlation reads $C_{2}(\tau
)\!=\!\left[ f(\mathbf{\hat{r}})(\hbar \omega _{A}/4\pi cr^{2})\right]
^{2}\!\gamma ^{2}\!\lim_{t\rightarrow \infty }\!\overline{\sigma ^{\dagger
}(t)\sigma ^{\dagger }(t+\tau )\sigma (t+\tau )\sigma (t)}.$ In these
expressions, $\gamma $ is the characteristic decay rate of the fluorescent
Markovian system.
\bibitem{despuez} The deduction and characterization of these phenomena in
the context of SMS will be published elsewhere.
\end{thebibliography}
\end{document}